\documentclass[numberedappendix]{emulateapj}
\usepackage{graphicx}
\usepackage{epstopdf}
\usepackage{natbib}
\newcommand{\beq}{\begin{equation}}
\newcommand{\eeq}{\end{equation}}
\newcommand{\beqa}{\begin{eqnarray}}
\newcommand{\eeqa}{\end{eqnarray}}

\newcommand{\simgt}{\lower.5ex\hbox{$\; \buildrel > \over \sim \;$}}
\newcommand{\simlt}{\lower.5ex\hbox{$\; \buildrel < \over \sim \;$}}

\def\imag{\dot \imath}

\def \sn2{\left(S/N\right)^2}

\begin{document} 

\title{New Neutrino Mass Bounds from Sloan Digital Sky Survey III Data Release 8 Photometric Luminous Galaxies}
\author{Roland de Putter\altaffilmark{1,2}, Olga Mena\altaffilmark{2}, Elena Giusarma\altaffilmark{2},
Shirley Ho\altaffilmark{3,4}, Antonio Cuesta\altaffilmark{5},
Hee-Jong Seo\altaffilmark{3,6}, Ashley J.~Ross\altaffilmark{7}, Martin White\altaffilmark{3,8},
Dmitry Bizyaev\altaffilmark{9}, Howard Brewington\altaffilmark{9},
David Kirkby\altaffilmark{10},
Elena Malanushenko\altaffilmark{9}, Viktor Malanushenko\altaffilmark{9},
Daniel Oravetz\altaffilmark{9}, Kaike Pan\altaffilmark{9}, Will J.~Percival\altaffilmark{7},
Nicholas P.~Ross\altaffilmark{3}, Donald P.~Schneider\altaffilmark{11,12},
Alaina Shelden\altaffilmark{9}, Audrey Simmons\altaffilmark{9},
Stephanie Snedden\altaffilmark{9}
}
\altaffiltext{1}{ICC, University of Barcelona (IEEC-UB), Marti i Franques 1, Barcelona 08028, Spain}
\altaffiltext{2}{Instituto de Fisica Corpuscular, University of Valencia-CSIC, Spain}
\altaffiltext{3}{Lawrence Berkeley National Laboratory, 1 Cyclotron Road, Berkeley, CA 94720}
\altaffiltext{4}{Carnegie Mellon University, Physics Department, 5000 Forbes Ave, Pittsburgh, PA 15213}
\altaffiltext{5}{Yale University, New Haven, CT}
\altaffiltext{6}{Berkeley Center for Cosmological Physics, University of California, Berkeley, CA 94720}
\altaffiltext{7}{Institute of Cosmology \& Gravitation, Dennis Sciama Building, University of Portsmouth, Portsmouth PO1 3FX, UK}
\altaffiltext{8}{Departments of Physics and Astronomy, University of California, Berkeley, CA 94720}
\altaffiltext{9}{Apache Point Observatory, P.O. Box 59, Sunspot, NM 88349-0059, USA}
\altaffiltext{10}{Department of Physics and Astronomy, University of California, Irvine, CA 92697}
\altaffiltext{11}{Department of Astronomy and Astrophysics, The Pennsylvania State University, University Park, PA 16802}
\altaffiltext{12}{Institute for Gravitation and the Cosmos, The Pennsylvania State University, University Park, PA 16802}


\date{\today}

\begin{abstract}
We present neutrino mass bounds using 900,000 luminous galaxies with photometric
redshifts measured from Sloan Digital Sky Survey III Data Release Eight (SDSS DR8). The galaxies have photometric redshifts between $z = 0.45$ and $z = 0.65$,
and cover 10,000 square degrees and thus probe a volume of 3$h^{-3}$Gpc$^3$,
enabling tight constraints to be derived on the amount of dark matter
in the form of massive neutrinos.
A new bound on the sum of neutrino masses $\sum m_\nu < 0.26$~eV, at 95\% confidence level (CL),
is obtained after combining our sample of galaxies, which we call "CMASS", with WMAP 7 year Cosmic Microwave Background (CMB)
data and the most recent measurement of the Hubble parameter from
the Hubble Space Telescope (HST).
This constraint is obtained with a conservative multipole range
choice of $30 < \ell < 200$ in order
to minimize non-linearities, and a free bias parameter in each of the four redshift bins.
We study the impact of assuming this linear galaxy bias model using mock catalogs,
and find that this model causes a small ($\sim 1-1.5 \sigma$) bias in $\Omega_{\rm DM} h^2$.
For this reason, we also quote neutrino bounds based on
a conservative galaxy bias model containing additional, shot noise-like free parameters.
In this conservative case, the bounds are significantly weakened, e.g.~$\sum m_\nu < 0.36$~eV (95\% confidence level)
for WMAP+HST+CMASS ($\ell_{\rm max}=200$).
We also study the dependence of the neutrino bound on multipole range ($\ell_{\rm max}=150$ vs $\ell_{\rm max}=200$)
and on which combination of data sets is included as a prior. The addition of supernova and/or Baryon Acoustic Oscillation data does
not significantly improve the neutrino mass bound once the HST prior is included.

A companion paper \citep{hoetal11} describes the construction of the angular power spectra in detail
and derives constraints on a general cosmological model, including the dark energy equation of state $w$
and the spatial curvature $\Omega_K$, while a second companion paper \cite{seobao} presents a measurement of the scale
of baryon acoustic oscillations from the same data set. All three works are based on the catalog by \cite{rossetal11}.
\end{abstract} 

\maketitle

\section{Introduction}
\label{sec:intro}

During the last several years, experiments involving solar, atmospheric, 
reactor and accelerator neutrinos have adduced robust evidence for flavor 
change, implying non-zero neutrino mass, see Ref.~\cite{GonzalezGarcia:2007ib} and references therein.
The most economical description 
of the neutrino oscillation phenomena requires at least two massive neutrino 
mass eigenstates to explain the two mass
differences\footnote{Neutrino oscillations are described by mass squared differences and
not by the absolute values of the mass eigenstates},
$\Delta m^2_{12}=7.59\cdot 10^{-5}$~eV$^2$ and $\Delta m^2_{23}=2.5\cdot 10^{-3}$~eV$^2$~\cite{Schwetz:2011zk,Fogli:2011qn},
which drive the solar and atmospheric 
transitions.
Despite the remarkable success 
of past and present oscillation experiments, and the promising prospects for 
future searches, the individual neutrino masses and the Dirac versus 
Majorana neutrino character are key questions that continue to be unanswered by oscillation experiments. 

Direct information on the absolute scale of neutrino masses 
can be extracted from kinematical studies of weak decays producing neutrinos. 
The present upper bound on the electron-neutrino mass from tritium beta-decay experiments
is $2$~eV ($95\%$ confidence level (CL))~\cite{Lobashev:2003kt,Eitel:2005hg}, and in the future the
KATRIN experiment is expected to be sensitive to electron-neutrino masses approaching $0.2$~eV ($90\%$ CL)~\cite{Otten:2008zz}.
Searches for the Majorana neutrino nature involve neutrinoless double beta decay $\beta\beta$($0\nu$),
a rare and as yet unobserved transition between two nuclei. Observational upper limits on $\beta\beta$($0\nu$)
rates provide an upper bound on the so-called “effective Majorana mass” of the electron neutrino,
$\langle m_{\rm eff}\rangle  < 0.3-1.0$~eV, bound which would only apply if neutrinos are Majorana particles~\cite{GomezCadenas:2011it}.
Forthcoming  $\beta\beta$($0\nu$) experiments aim for sensitivity approaching $\langle m_{\rm eff}\rangle  < 0.05$~eV~\cite{GomezCadenas:2011it}.

Cosmology provides one of the means to tackle the absolute scale of neutrino 
masses. Some of the earliest cosmological bounds on neutrino masses followed 
from the requirement that massive relic neutrinos, present today in the 
expected numbers, do not saturate the critical density of the Universe, 
i.e., that the neutrino mass energy density given by
\begin{equation}
\label{eq:omnu}
\Omega_\nu=\frac{\sum m_\nu}{93.1 h^2 \rm{eV}}
\end{equation}
satisfies $\Omega_\nu\le 1$. The Universe therefore offers a new laboratory 
for testing neutrino masses and neutrino physics. Accurate 
measurements of the Cosmic Microwave Background (CMB) temperature and polarization anisotropy 
from satellite, balloon-borne and ground-based experiments 
have fully confirmed the predictions of the standard cosmological model and 
allow us to weigh neutrinos~\cite{Lesgourgues:2006nd}. Indeed, neutrinos can play a relevant 
role in large-scale structure formation and leave key signatures in 
several cosmological data sets. More specifically, the amount of primordial 
relativistic neutrinos changes the epoch of matter-radiation equality, 
leaving an imprint on CMB anisotropies. After becoming non-relativistic, their free-streaming nature
damps power on small scales, suppressing the growth of matter density 
fluctuations and thus affecting both the CMB
and galaxy clustering observables in the low-redshift universe~\cite{Lesgourgues:2006nd}.
Measurements of all of these observations have been used to place new bounds on neutrino physics from
cosmology~\cite{Elgaroy:2002bi,Spergel:2003cb,Hannestad:2003xv,Allen:2003pta,Tegmark:2003ud,Barger:2003vs,Hannestad:2003ye,Crotty:2004gm,Seljak:2004xh,Elgaroy:2004rc,Hannestad:2005gj,Goobar:2006xz,Spergel:2006hy,Seljak:2006bg,Fogli:2008ig,Komatsu:2008hk,Reid:2009xm,Reid:2009nq,Thomas:2009ae,reidetal10,Komatsu:2010fb,saitoetal11,riemeretal11,bensonetal11}, with a current
limit on the sum of neutrino masses
$\Sigma m_{\nu} \simlt 0.6$~eV
at $95\%$ CL (e.g.~\cite{Reid:2009xm}),
depending on the precise combination of data sets and on the underlying cosmological model. 

We present here neutrino mass bounds from the final imaging data set of the Sloan Digital Sky Survey (SDSS-III)~\cite{SDSS}, using the photometric redshift catalog of \emph{Ross et al.}~\cite{rossetal11}.
We consider the CMASS sample~\cite{whiteetal11} of luminous galaxies of SDSS DR8~\cite{DR8}, the eighth data release of SDSS and the first
data release of the Baryon Oscillation
Spectroscopic Survey (BOSS) \cite{eisetal11}, with photometric
redshifts $z=0.45 - 0.65$. This sample covers an area of approximately 10,000 square degrees and consists of 900,000 galaxies. It is thus the largest sample of luminous galaxies so far and promises strong constraints on the neutrino properties (see \cite{Thomas:2009ae} for an analysis of a slightly smaller SDSS photometric sample).

We derive neutrino constraints from the angular power spectra
of the galaxy density at different redshifts, in combination with priors from the CMB
and from measurements of the Hubble parameter, supernovae distances and
the BAO scale. The spectra and the analysis of a minimal $\Lambda$CDM
cosmology are described in detail in our companion paper
\cite{hoetal11} and the measurement of the BAO scale from the spectra is 
presented in a separate companion paper \cite{seobao}.
We will often refer to these works for details and focus here on the neutrino bound.

The structure of the paper is as follows. In section \ref{sec:data}, we describe the data set and
the derived angular spectra. We then discuss our theoretical model for the spectra
and their cosmology dependence in section \ref{sec:model}. In section \ref{sec:nu} we explain
the specific signature of neutrino mass on galaxy clustering data.
We test our model for the angular power spectra against mock data in
section \ref{sec:mocks}
and present the constraints on the sum of the neutrino
masses and other parameters for several data combinations in section \ref{sec:results}.
Finally, we discuss these results and conclude in section \ref{sec:disc}.

\section{Data}
\label{sec:data}

The data and the method for obtaining angular spectra have been described 
in detail in Ref.~\cite{rossetal11} and in \cite{hoetal11}. Here we 
provide a brief description of the main properties and refer the reader 
to those papers for details. Our galaxy sample is obtained from imaging 
data from DR8 \cite{DR8} of SDSS-III~\cite{SDSS}. This survey mapped about 
$15,000$ square degrees of the sky in five pass bands ($u, g, r, i$ and $z$)~\cite{fukugitaetal96}
using a wide field CCD camera \cite{gunnetal98} on the $2.5$ meter Sloan telescope
at Apache Point Observatory \cite{gunnetal06} (the subsequent astrometric calibration
of these imaging data is described in \cite{pieretal03}). A sample of $112,778$ galaxy spectra from 
BOSS~\cite{eisetal11} 
were used as a training sample for the photometric redshift catalog, as described 
in \cite{rossetal11}.

\begin{figure*}[!htb]
\centering
  \includegraphics*[width=\linewidth]{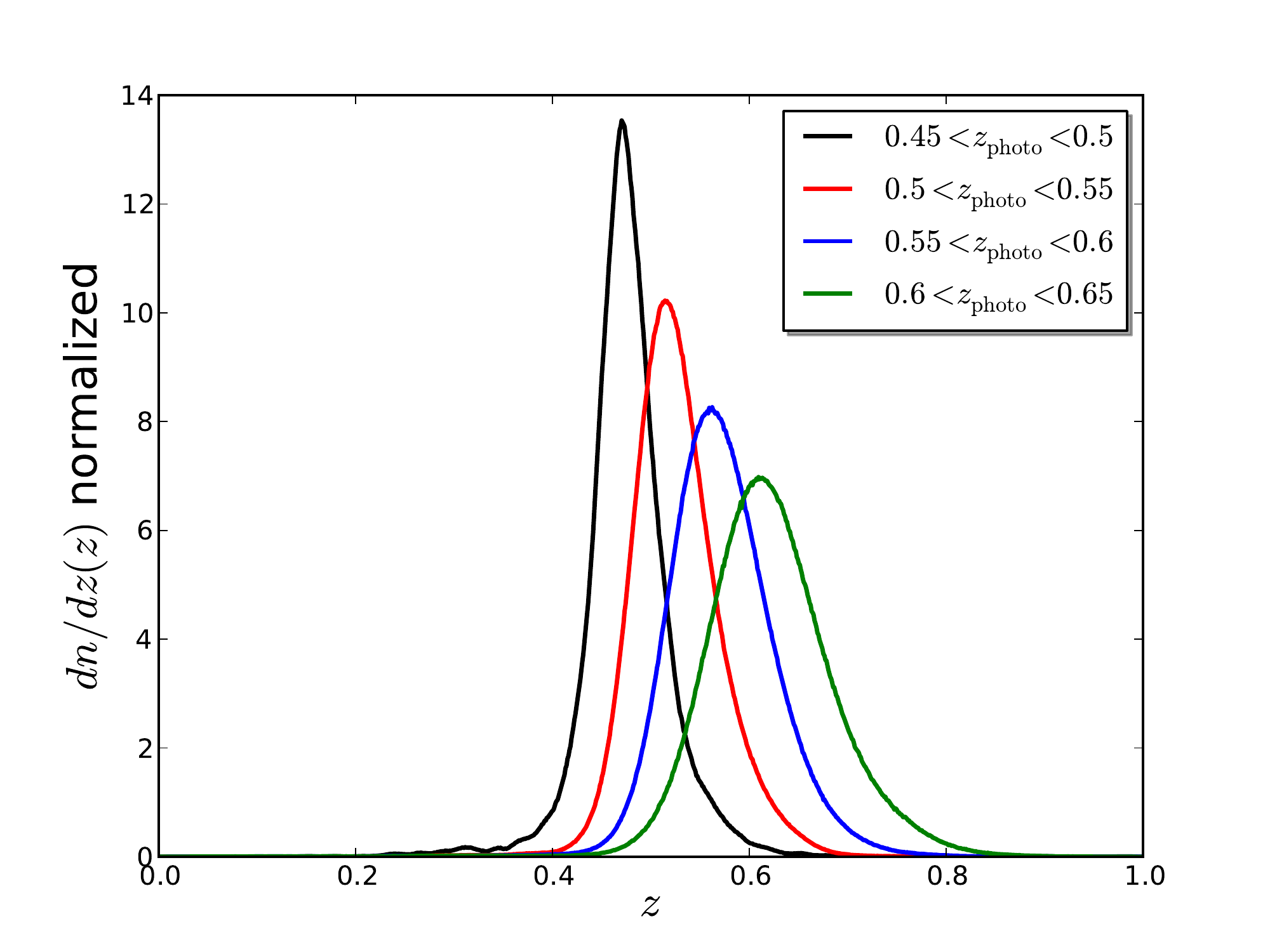}
  \caption{Normalized true redshift distribution of CMASS galaxies in four photometric redshift bins. The number of galaxies
  in each bin is $214971$, $258736$, $248895$ and $150319$ (from low to high redshift).
}
  \label{fig:dndz}
\end{figure*}

\begin{figure*}[!htb]
\begin{minipage}[t]{0.49\textwidth}
\centering
  \includegraphics*[width=\linewidth]{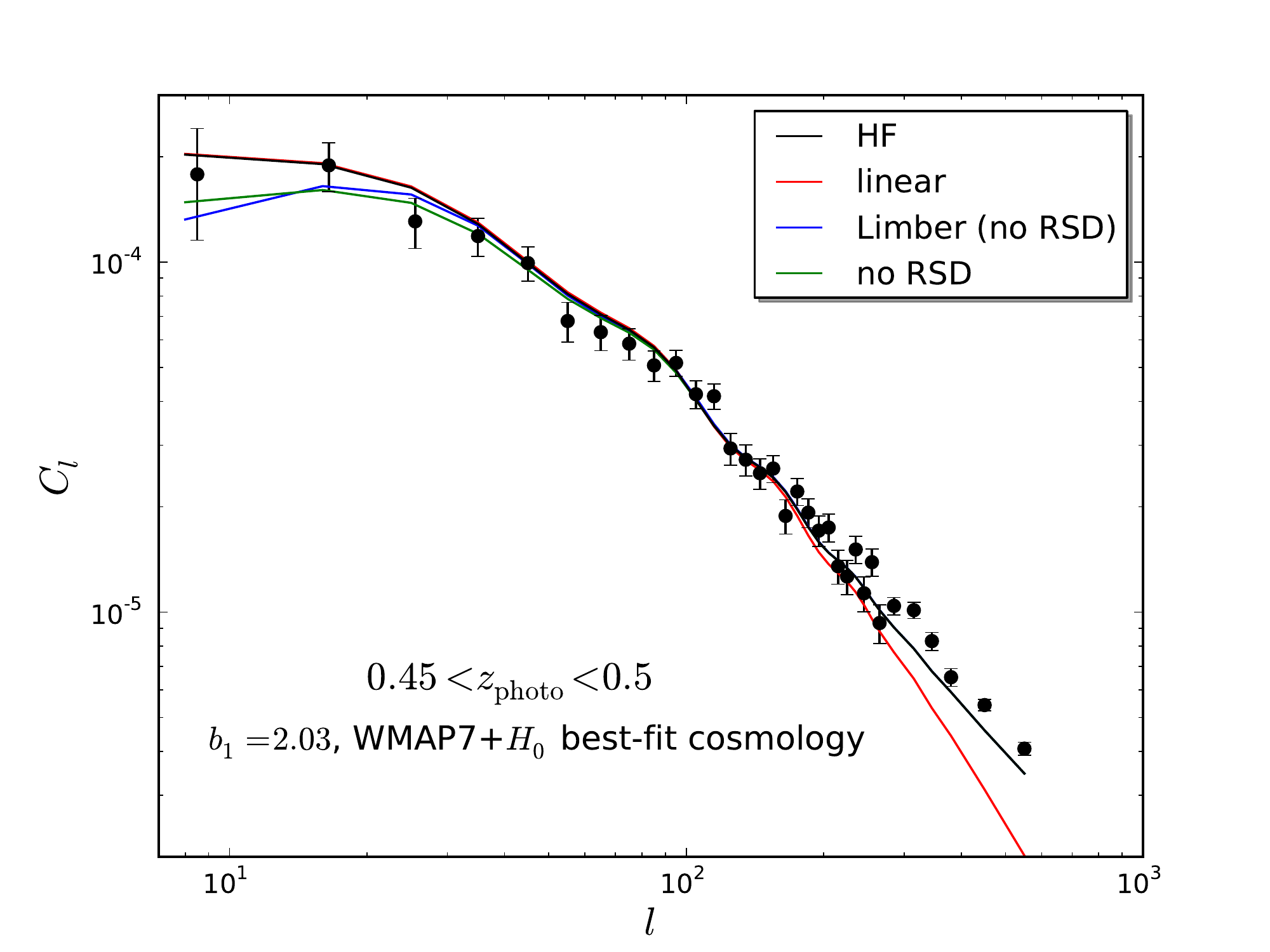}
  \includegraphics*[width=\linewidth]{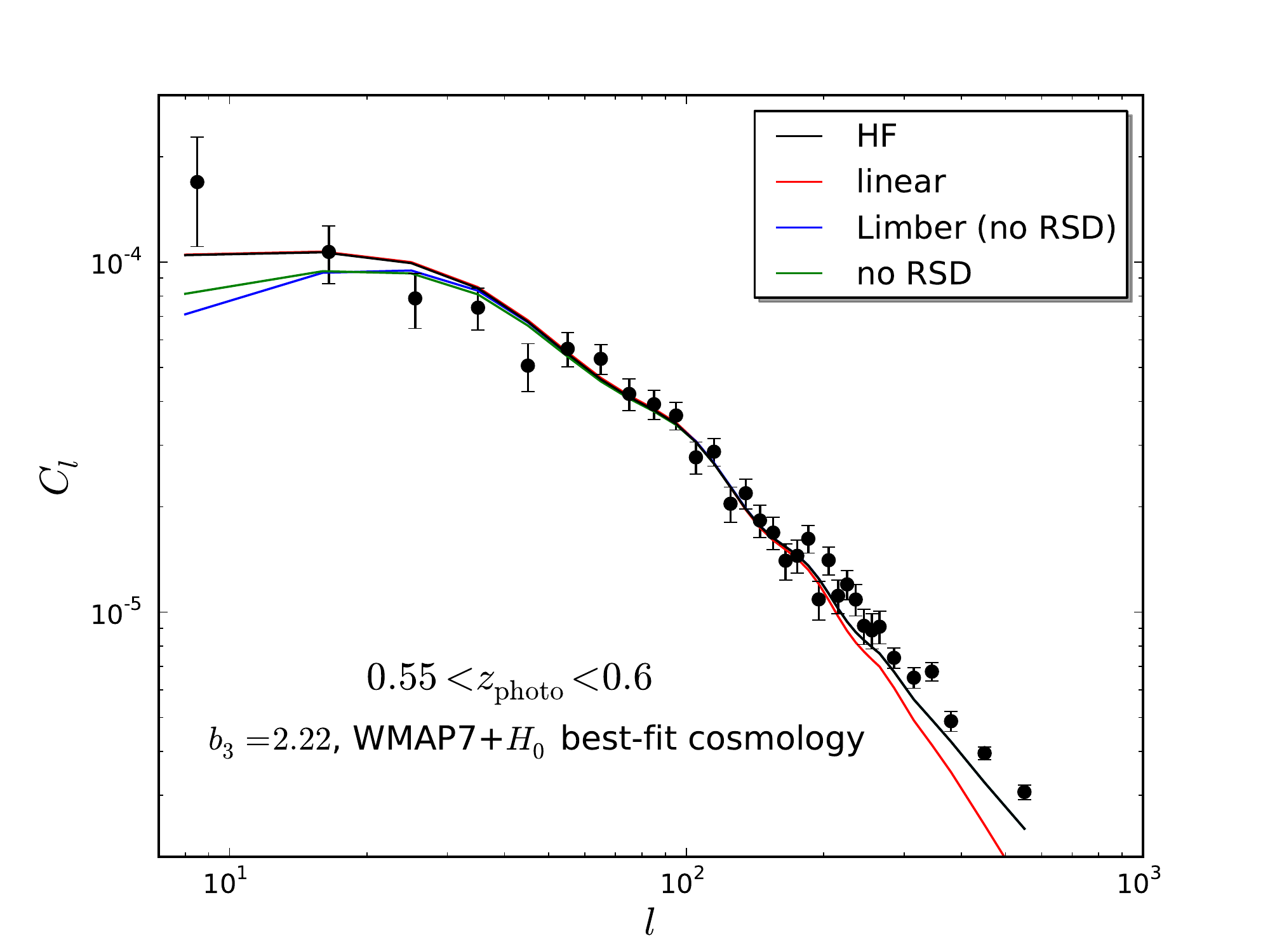}
\end{minipage} \hfill
\begin{minipage}[t]{0.49\textwidth}
\centering
  \includegraphics*[width=\linewidth]{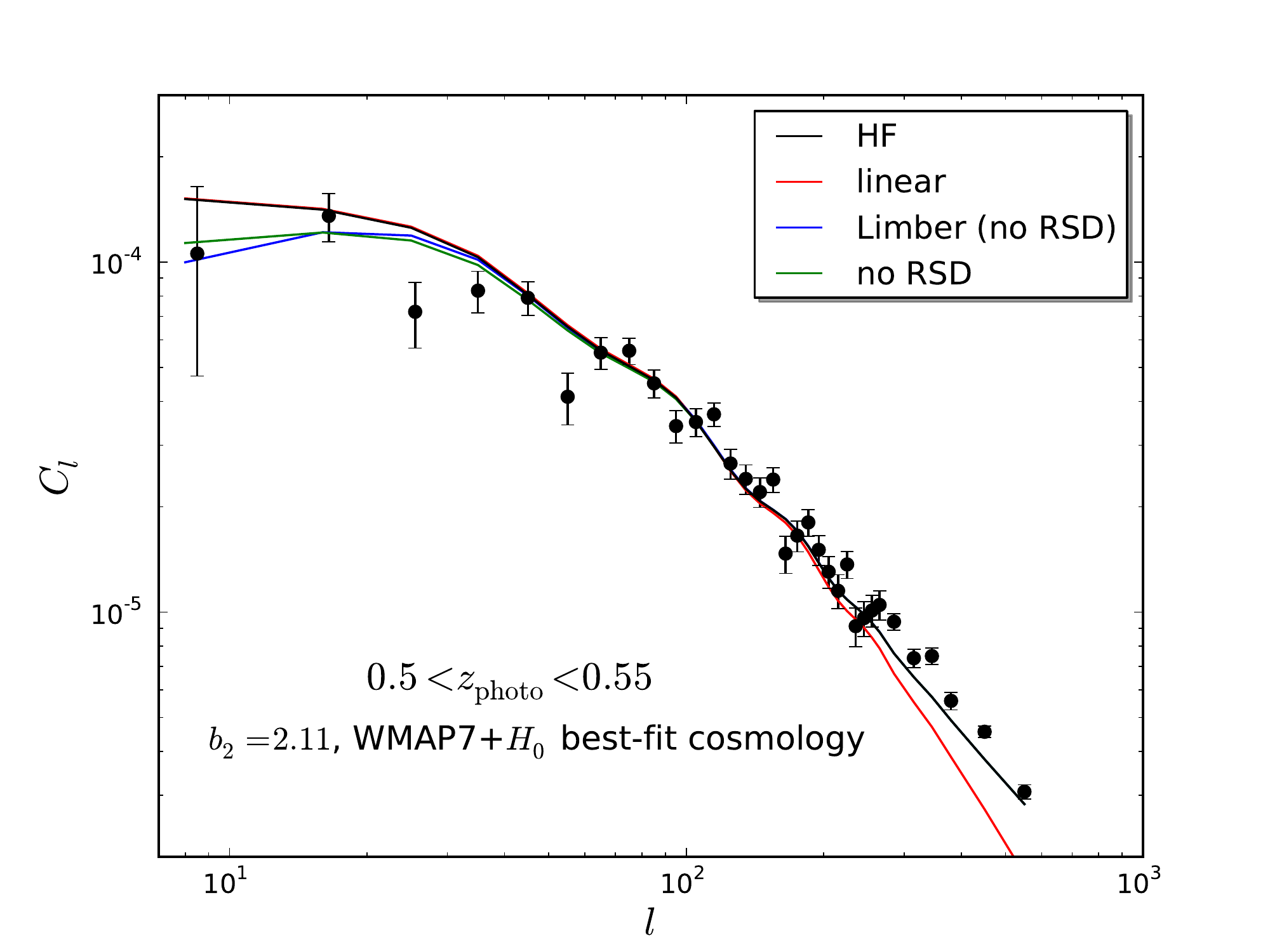}
  \includegraphics*[width=\linewidth]{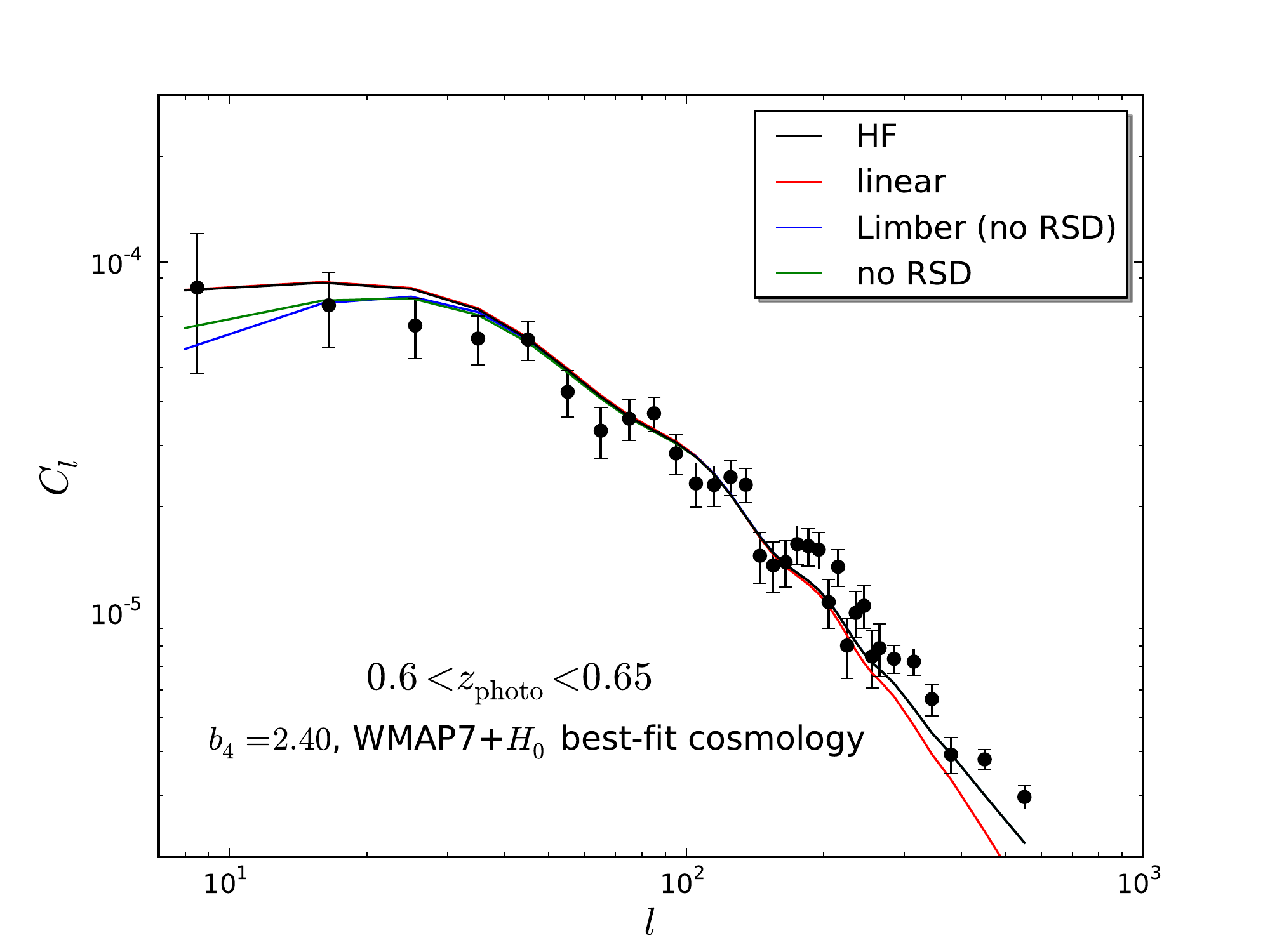}
\end{minipage} \hfill
  \caption{Observed power spectra (black points) with error bars and theoretical power
spectra (solid curves).  We show the theoretical power spectra for different models: 
the default, HaloFit (HF) based model used in our analysis (black; see text for details),
the same model, but using the linear matter power spectrum as input (red),
  the default model, but using the Limber approximation (blue) and the default model without redshift space distortions (green).
  We restrict ourselves to the range $\ell = 30 - 200$ in our analysis. For the theoretical
  spectra, we assume the WMAP7+HST best-fit cosmology and use the bias $b_i$ that best fits the data.
  We do not here include the shot noise parameters $a_i$.
}
  \label{fig: data}
\end{figure*}

We focus on the approximately stellar mass-limited CMASS sample of luminous 
galaxies, detailed in \cite{whiteetal11}, which are divided into four 
photometric redshift bins, $z_{\rm photo}=0.45-0.5-0.55-0.6-0.65$. 
The photometric redshift error lies in the range $\sigma_z(z) = 0.04 - 0.06$,
increasing from low to high redshift. Figure \ref{fig:dndz} shows
the estimated true redshift distribution of each bin,
determined using the methods described in section 5.3 of \cite{rossetal11}.

The calculation of the angular power spectrum for each bin 
is described in detail in our companion paper \cite{hoetal11} and uses the optimal quadratic 
estimator (OQE) method outlined in \cite{seljak98,tegmarketal98,padmaetal03,padmaetal07}.
The four power spectra are binned in $\ell$ space with a typical 
wave band width of $\Delta \ell = 10$.
The expectation value of the power spectrum in a wave band is a convolution of
the true power spectrum with a window function of width roughly equal to the 
typical wave band width. Examples of these window functions are shown in 
Fig.~3 of the companion paper \cite{seobao}.
When fitting the data to the underlying theoretical model, 
we always apply these window functions to the theoretical power
spectra before calculating the likelihood relative to the data.

The four power spectra are plotted with their error bars in 
Fig.~\ref{fig: data}. The solid curves represent theoretical power spectra based on
several assumptions. These power spectra will be discussed in detail in section \ref{sec:model}.
Since the low $\ell$ wave bands are more prone to systematics~\cite{rossetal11},
we are conservative and do not consider bands with $\ell < 30$ in our analysis.
We shall apply cuts at $\ell_{\rm max} = 150$ and $200$ in order to suppress uncertainties from
non-linear corrections to the modeled power spectra, as discussed in the following sections.
The median redshift ($z \approx 0.55$) contributions to these maximum angular modes
arise from three-dimensional modes with wave vectors $k\approx 0.10 h$Mpc$^{-1}$ and $k\approx 0.14 h$Mpc$^{-1}$,
respectively. We thus use 17 (12) data points per redshift slice for $\ell_{\rm max} = 200 \, (150)$.

\section{Modeling the angular power spectra}
\label{sec:model}

The galaxy overdensity in the $i$-th redshift bin can be expanded in terms of spherical harmonics,
\beq
\delta_g^{(i)}(\hat{n}) = \sum_{\ell m} a_{\ell m}^{(i)} \, Y_{\ell m}(\hat{n})~,
\eeq
so that the angular power (and cross) spectra are defined as
\beq
\langle a_{~\ell m}^{(i)} a_{\ell'm'}^{(j)*} \rangle \equiv C^{(ij)}_\ell \, \delta^{\rm K}_{\ell \ell'}\,\delta^{\rm K}_{mm'},
\eeq
where $\delta^{\rm K}_{ij}$ is the Kronecker delta function.
As mentioned in the previous section, we do not estimate our spectra by directly transforming the observed density field to harmonics space,
but use instead the optimal quadratic estimator technique. To constrain the sum of the
neutrino masses
and other cosmological parameters, the observed spectra are compared to a cosmology dependent
model, which we now describe.

The galaxy overdensity on the sky is
a line-of-sight projection of the three-dimensional redshift space galaxy overdensity
$\delta_g (d(z) \, {\bf \hat{n}}, z)$,
\beq
\delta_g^{(i)}({\bf \hat{n}}) = \int dz \, g_i(z) \, \left(\delta_g(d(z) \, {\bf \hat{n}}, z) - (H(z))^{-1}\, {\bf \hat{n}} \cdot {\bf \nabla} ({\bf \hat{n}} \cdot {\bf v}(d(z) \, {\bf \hat{n}}, z))\right),
\label{eq:proj}
\eeq
where
\beq
g_i(z) = \frac{dn_i/dz(z)}{\int dz' \, dn_i/dz(z')}
\eeq
is the normalized redshift distribution of galaxies in bin $i$ (with $dn_i/dz(z)$
the number of galaxies per steradian per unit redshift), $d(z)$
is the comoving distance to redshift $z$ (assuming a flat universe) and ${\bf v}$ is the galaxy velocity field.
The velocity term arises because gradients of the
peculiar velocity contribution to the distance in redshift space change the volume, and consequently,
the number density\footnote{Instead of writing the projected galaxy overdensity as an integral
over the observed redshift (including peculiar velocity contributions) as in Eq.~(\ref{eq:proj}),
one could equivalently do the integral over ``true'' cosmic redshift, see Ref.~\cite{padmaetal07},
in which case only the true three-dimensional galaxy overdensity appears explicitly and the redshift
space distortions come in through a modification of the distribution $g_i(z)$.}.

We assume a linear, scale-independent bias for the galaxy density,
\beq
\delta_g({\bf x}, z) =  b_g(z) \, \delta_m({\bf x}, z)~,
\eeq
with $\delta_m$ the matter overdensity. For the peculiar velocity field, we use the continuity equation in the linear regime,
which gives for a Fourier mode with wave vector ${\bf k}$,
\beq
{\bf v} = -\imag \beta(z) \delta_g({\bf k}) \frac{{\bf k}}{k^2}~,
\eeq
where $\beta(z) = f(z)/b_g(z)$ is the redshift distortion parameter and
\beq
f(z) \equiv \frac{d\ln D(z)}{d\ln a}
\eeq
is the growth factor (with $D(z)$ the linear growth function).
 In the presence of neutrinos, the growth function is no longer scale 
 independent at late time as the neutrinos suppress growth on scales 
 below the free streaming length~\cite{Hu:1997vi,Eisenstein:1997jh}, but not on larger scales 
 (with a broad transition regime in between). We shall ignore the scale 
dependent growth in $\beta(z)$ since it is a small ($\ll 10 \%$) correction to the already 
 small effect (on the scales of interest here) of redshift space distortions (RSD, see Fig.~\ref{fig: data}). 
  However, this scale-dependent growth is included in the real space 
 power spectrum, as this is the main signature of massive neutrinos.

We simplify our treatment of the galaxy bias by following the approach of our two companion papers, adding four free
parameters $b_i$ to describe the bias in each bin. The results from our simulations barely change when considering a
bias $b_g(z)$ that varies within redshift bins, showing that this is a safe approximation\footnote{A similar approach is
considered to model $\beta(z)$, appearing in the redshift space distortion contribution.
For each bin we calculate an effective growth rate $f_i = \left(\Omega_{\rm DM}(z_i)\right)^{0.56}$
where $z_i$ is the mean redshift of the $i$-th bin, ignoring the scale dependence of the RSD growth.}.
It then follows from the above (see \cite{fisheretal94,heavenstaylor95,padmaetal07}) 
that 
\beq
\label{eq:clfull}
C_\ell^{(ii)} = b_i^2 \, \frac{2}{\pi} \int k^2 dk \, P_m(k,z=0) \, \left( \Delta_\ell^{(i)}(k) + \Delta^{{\rm RSD}, (i)}_\ell(k) \right)^2,
\eeq
where $P_m(k,z=0)$ is the matter power spectrum at redshift zero and
\beq
\label{eq:leg}
\Delta_\ell^{(i)}(k) = \int dz \, g_i(z) \, T(k, z) \, j_\ell(k \, d(z))~.
\eeq
Here, $j_\ell$ is the spherical Bessel function and $T(k, z)$ 
the matter transfer function
relative to redshift
zero\footnote{The transfer function is defined as $\delta_m({\bf k}, z) = T(k, z) \, \delta_m({\bf k}, z=0)$.}.
The contribution due to redshift space distortions is
\beqa
\label{eq:legrsd}
\Delta^{{\rm RSD}, (i)}_l(k) &=& \beta_i \, \int dz \, g_i(z) \, T(k, z) \, 
   \left[\frac{(2l^2+2l-1)}{(2l+3)(2l-1)} j_l(kd(z)) \right. \nonumber \\
   &&\left.- \frac{l(l-1)}{(2l-1)(2l+1)} j_{l-2}(kd(z)) \right. \nonumber \\
   &&\left.- \frac{(l+1)(l+2)}{(2l+1)(2l+3)} j_{l + 2}(k d(z)) \right]~.
\eeqa

To compute the matter power spectrum at a given redshift $P_m(k, z) = P_m(k,z=0)\, T^2(k, z)$,
we first make use of the CAMB code \cite{LewChalLas00},
which provides the linear power spectrum by integrating  the Boltzmann equations of all species including massive neutrinos.
We then apply the HaloFit
prescription\footnote{Recently, \cite{birdetal11} developed an extension to HaloFit
that incorporates the effect of massive neutrinos. We do not use this prescription as the correction to
standard HaloFit is negligible on the scales of our interest.}  \cite{Smithetal03} to the linear power spectrum
to account for non-linear effects on the matter power spectrum. 

While in the linear regime the galaxy spectrum is easy to model, calculations on non-linear scales inevitably 
have large uncertainties.
This effect is aggravated by the presence of 
massive neutrinos since for the massive neutrino case the non-linear regime 
has been explored less extensively in the literature than for a vanilla $\Lambda$CDM model.
In the non-linear regime, the matter power spectrum receives corrections due to gravitational collapse,
the galaxy bias becomes scale-dependent, and
redshift space distortions receive important contributions from velocity dispersion in collapsed objects. We
take into account non-linear corrections to the matter spectrum using HaloFit. The effect of non-linearities on
redshift space distortions
at the relevant scales here is small as it is largely washed out by line-of-sight
projection. However, we do expect significant corrections
to our model on small scales due to non-linear galaxy bias,
which we address below.

\begin{figure*}[!htb]
\begin{minipage}[t]{0.49\textwidth}
\centering
  \includegraphics*[width=\linewidth]{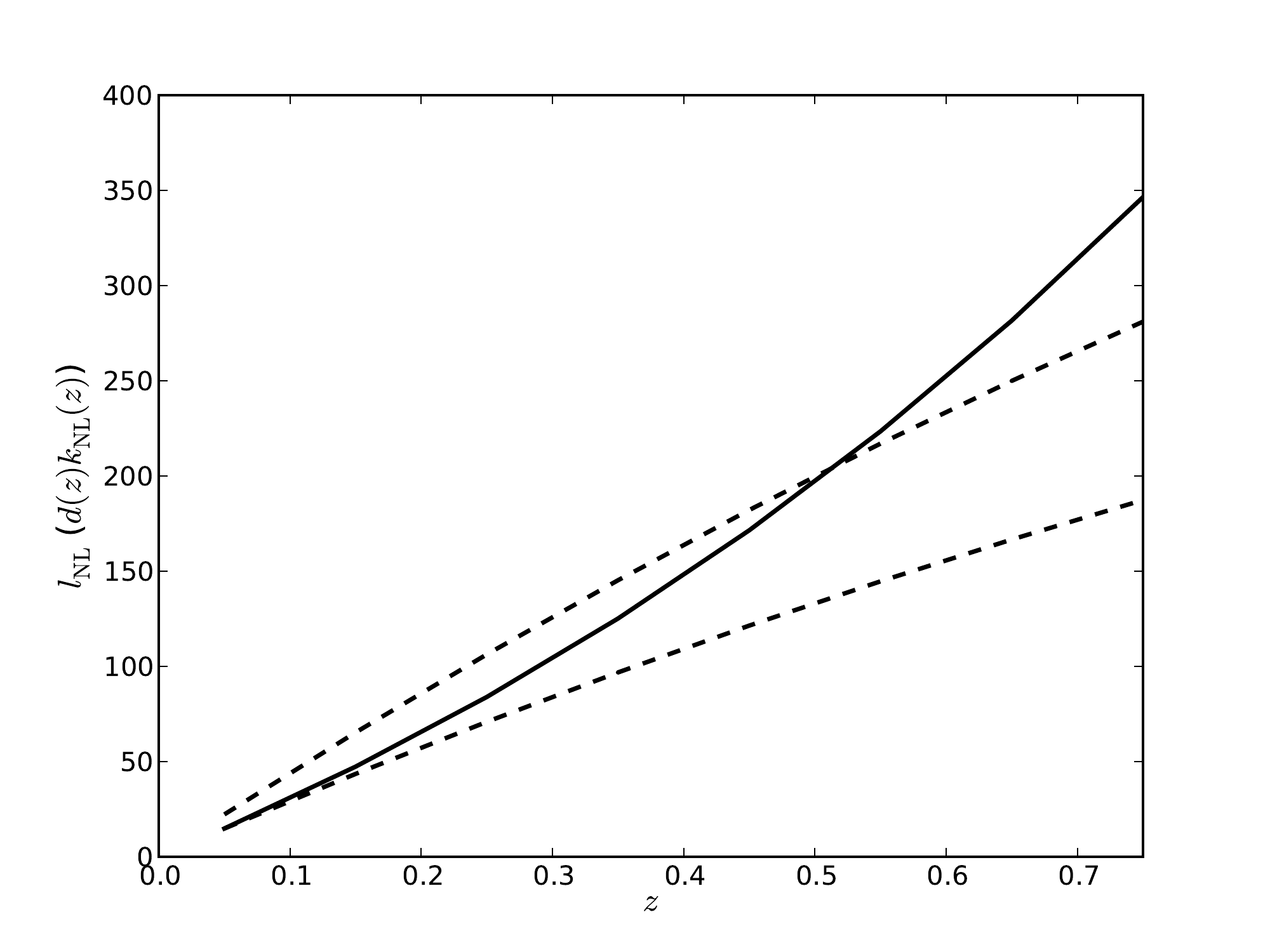}
\end{minipage} \hfill
\begin{minipage}[t]{0.49\textwidth}
\centering
  \includegraphics*[width=\linewidth]{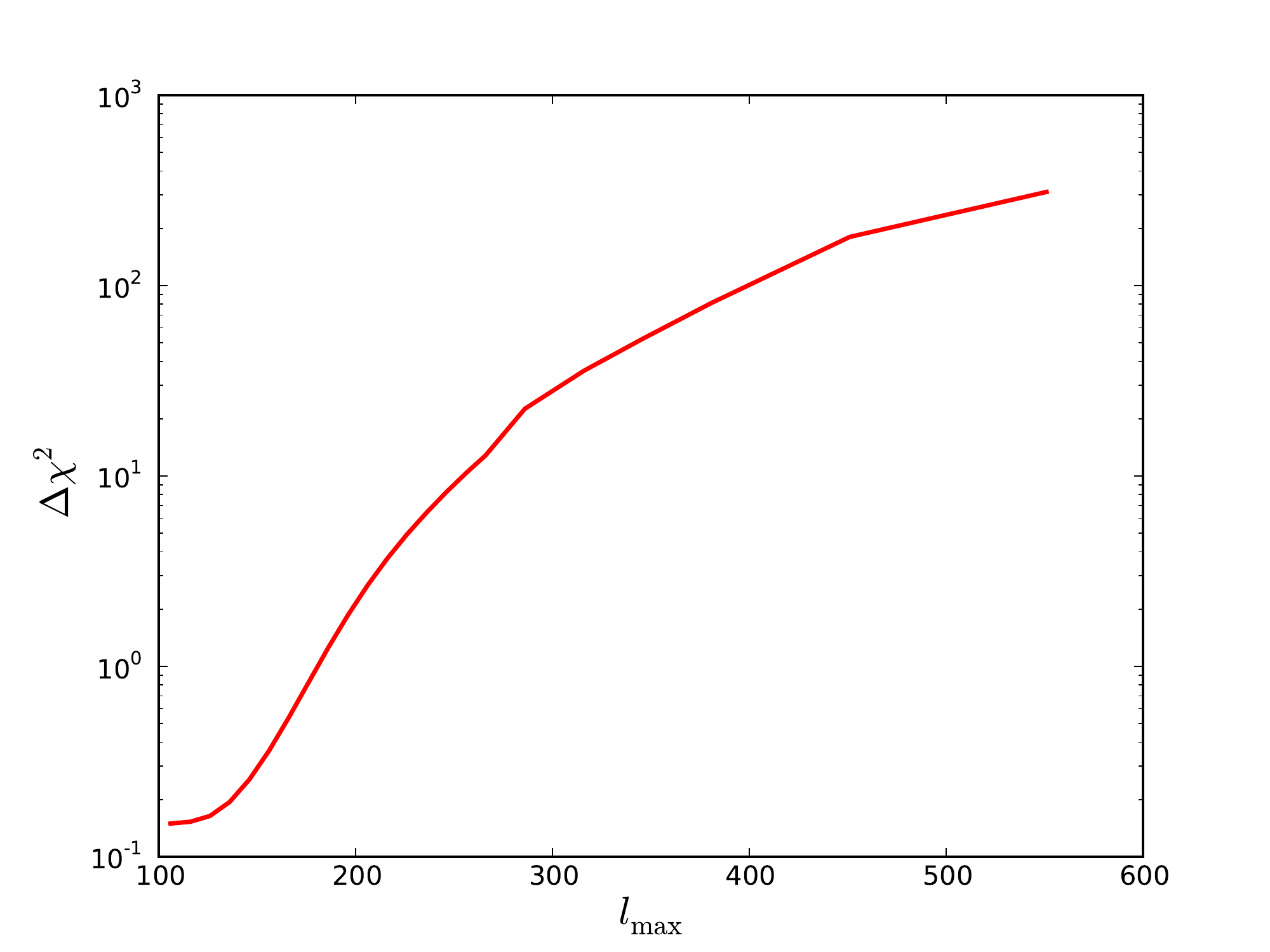}
\end{minipage} \hfill
  \caption{{\em Left panel}: Minimum multipole at which 3-D power spectrum contribution to the angular power spectrum
  receives important non-linear corrections, as
  a function of redshift, $\ell_{\rm NL} \equiv k_{\rm NL} \, d(z)$. We consider several choices
  of the non-linear scale $k_{\rm NL}$. The dashed curves are for
  $k = 0.15 h$Mpc$^{-1}$ (top) and $0.1 h$Mpc$^{-1}$ (bottom). The solid curve is for a simple
  model of a redshift dependent $k_{\rm NL}(z) = R_{\rm NL}(z=0)/R_{\rm NL}(z) \times 0.1 h$Mpc$^{-1}$,
  where $R_{\rm NL}(z)$ is such that the matter overdensity variance averaged over spheres with this radius equals one
  (using the linear power spectrum).
{\em Right panel}: The $\chi^2$ difference as a function of $\ell_{\rm max}$
between our default template, which uses Halofit, and a template using the linear matter power spectrum,
given the covariance matrix for the CMASS spectra.
We assume the WMAP7 plus HST best fit cosmology and fix the bias parameters $b_i = 2$ ($a_i=0$).
Both plots suggest that non-linear effects start to become (mildly) relevant at $\ell_{\rm max}$
between 150 and 200.
}
  \label{fig: ell_nl}
\end{figure*}

For angular scales where non-linear effects cannot be ignored, the contribution
to a given angular mode $\ell$ from a redshift $z$ comes exclusively 
from three-dimensional modes with wave vector $k\approx \ell/d(z)$.
Clearly, to avoid large non-linear corrections, the analysis must be restricted to low $\ell$.
On the other hand, the density of modes per unit $\ell$ 
increases with $\ell$ so we want to use as many modes as possible without biasing the results.
 Figure~\ref{fig: ell_nl} (left panel) depicts (as a function of redshift $z$) the value of $\ell$ above which
 non-linear corrections to the three-dimensional power spectrum contributions to
 the angular spectrum become important (i.e.~$\ell_{\rm NL} \equiv k_{\rm NL}(z) \, d(z)$),
 considering various assumptions for the non-linear scale $k_{\rm NL}(z)$.
 Given that most of our signal is produced in the range $z=0.45-0.65$, and assuming that
 our model becomes inadequate at $k > 0.15 h$Mpc$^{-1}$, we conclude that a conservative choice would be
 $\ell_{\rm max}$ somewhere between 150 and 200.
 
 Alternatively, we can obtain an indication of the importance of non-linear galaxy bias
 by considering the effect of non-linear corrections to the {\it matter} power
 spectrum\footnote{However, one must keep in mind that this may underestimate
 the effect of non-linear galaxy bias, as galaxies are more strongly clustered than matter and are thus
 prone to larger non-linear corrections.} (which we {\it do}
 include in our model).
 The right panel in Figure \ref{fig: ell_nl} therefore shows
 the signal to noise ratio squared in the difference between our default model and
 the same model, but using the linear matter power spectrum instead of the
 non-linear (HaloFit) one. The signal to noise reaches one somewhere between
 $\ell_{\rm max} = 150$ and $200$, corresponding to contributions
 of modes $k_{\rm max}\approx 0.10 h$Mpc$^{-1}$ and $k_{\rm max}\approx 0.14 h$Mpc$^{-1}$
 at the median redshift $z=0.55$.
 Finally, a more concrete indication of the importance of non-linear galaxy bias
 to the range of scales of our choice is given by Fig.~13 (left panel) of \cite{hamausetal10},
 which shows the halo bias as a function of three-dimensional mode $k$. Since for our sample
 of galaxies the bias $b \sim 2$, the plot confirms that there is only a mild bias variation
 in the relevant range of three-dimensional scales relevant to the multipole range we have chosen.
 
 Based on the above discussion, we choose a default value $\ell_{\rm max} = 200$, but we will also
 present results for the more conservative choice $\ell_{\rm max} = 150$.
 While it is possible to model the galaxy spectrum in a more sophisticated
 manner (see e.g.~\cite{saitoetal08,saitoetal09,saitoetal11} for an approach based on perturbation
 theory and the local bias model \cite{mcdonald06}, and \cite{swansonetal10} for a cross-comparison of a number of methods),
 we consider it appropriate, given the multipole range we include, to
 use the simple model described in Eq.~(\ref{eq:clfull}), characterized by bias parameters $b_i$.
 In addition to this model,
we also consider an alternative model with more freedom, by adding shot noise-like parameters $a_i$
(see also our companion papers),
\beq
\label{eq:clfull2}
C_\ell^{(ii)} = b_i^2 \, \frac{2}{\pi} \int k^2 dk \, P_m(k,z=0) \, \left( \Delta_\ell^{(i)}(k) + \Delta^{{\rm RSD}, (i)}_\ell(k) \right)^2 + a_i~.
\eeq
The parameters $a_i$ serve to mimic effects of scale-dependent
galaxy bias
and to model the effect of potential insufficient
shot noise subtraction.
This model is a version of what is sometimes referred to as the ``P-model'' (e.g.~\cite{hamannetal08, swansonetal10})
and is independently motivated by the halo model \cite{seljak2000,seljak01,schulzwhite06,guziketal07} and the local bias ansatz \cite{scherrwein98,colesetal99,saitoetal09}.
We further discuss the validity of the parameterizations with and without $a_i$
in section \ref{sec:mocks}.

Figure~\ref{fig: data} shows the theoretical galaxy spectra as described in this section. The error bars follow from
the optimal quadratic estimator method used to construct the power spectra (see \cite{hoetal11} for details).
Comparing the spectra with
(black) and without (green) redshift space distortions shows that this effect is negligible for $\ell > 50$ and is
probably not relevant for the range of scales we use in our data analysis, i.e.~$\ell > 30$.
Although we never employ it, we also show the effect of using the Limber approximation \cite{Limber:54} and
find that for $\ell > 30$ it works excellently.

\section{Cosmological Signature of Neutrinos}
\label{sec:nu}

In the analysis presented in this paper we assume that there are three species of massive
neutrinos with equal masses $m_\nu$. Massive neutrinos affect galaxy formation at scales
below the Hubble horizon when they become non relativistic,
\begin{equation}
k_{nr} \simeq 0.0145 \left(\frac{m_\nu}{1 \ \textrm{eV}}\right) \sqrt{\Omega_{\rm DM}} \ h \textrm{Mpc}^{-1}~,
\end{equation}
with $\Omega_{\rm DM}$ the present total dark matter energy density, i.e.~cold dark matter plus massive neutrinos,
relative to the critical density.
The non-relativistic neutrino overdensities cluster at a given redshift $z$ only at scales where the wavenumber of perturbations is below the neutrino free streaming scale 
\begin{equation}
k_{fs}(z) = \frac{0.677}{(1+z)^{1/2}} \left(\frac{m_\nu}{1 \ \textrm{eV}}\right) \sqrt{\Omega_{\rm DM}} \ h \textrm{Mpc}^{-1}~,
\end{equation}
due to the pressure gradient, which prevents gravitational clustering.  On spatial scales larger than the
free streaming scale $k<k_{fs}$, neutrinos cluster and behave as cold dark matter (and baryons).
Perturbations with comoving wavenumber larger than the free streaming scale can not grow due to the
large neutrino velocity dispersion. As a consequence, the growth rate of density perturbations decreases and
the matter power spectrum is suppressed at $k>k_{fs}$. Since the free streaming scale depends on the individual neutrino masses and not on
their sum, a measurement of $k_{fs}$ could, in principle, provide the ordering of the
neutrino mass spectrum. In practice, such a task appears extremely challenging, see \cite{Jimenez:2010ev}. 

\begin{figure*}[!htb]
\centering
  \includegraphics*[width=\linewidth]{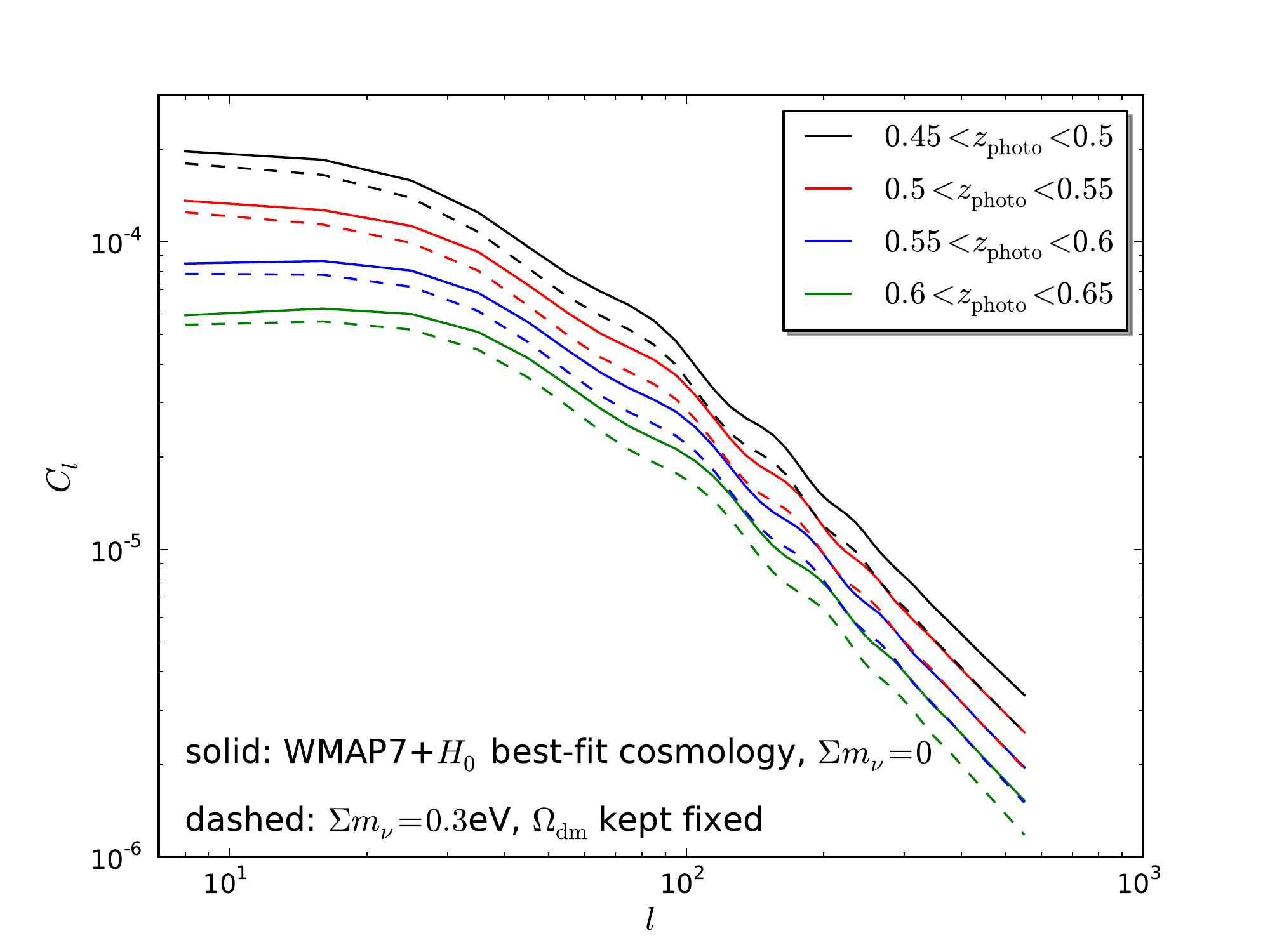}
  \caption{Effect of neutrinos on the angular power spectra. The solid and dashed curves depict the massless and $\Sigma m_{\nu} = 0.3$~eV cases, respectively.}
  \label{fig:nueffect}
\end{figure*}

Figure~\ref{fig:nueffect} illustrates the effect of massive neutrinos on the angular power spectra. The solid curves depict the results for the four redshift bins exploited here in the case of a $\Lambda$CDM model assuming no massive neutrino species and best fit parameters to WMAP7 year data~\cite{Larsonetal10,Komatsuetal10} and HST-$H_0$ data~\cite{hst11}. The dashed curves denote the angular power spectra results assuming three massive neutrinos with $\sum m_\nu=0.3$~eV and keeping the cold dark matter mass energy density constant.
In the presence of massive neutrinos the angular power spectra are suppressed
at each redshift at an angular scale that is related to the free streaming scale by $\ell \sim d(z) \, k_{fs}(z)$.
Therefore, the larger the neutrino mass (or the redshift), the larger the lowest angular wavenumber
at which the power spectrum is maximally suppressed.  In the redshift range of interest here and for
$\sum m_\nu=0.3$~eV, the suppression angular scale appears in the range $\ell=20-50$ (however, there is some suppression
even at lower $\ell$). Note as well that
there will exist a strong degeneracy between the neutrino masses and the amount of cold dark matter, since,
in principle, one could partially compensate the growth suppression induced by massive neutrinos at scales $k>k_{fs}$
by increasing the cold dark matter mass-energy density.
Combination with CMB and $H_0$ data will help to break this degeneracy. 

Neutrino masses affect the angular power spectra $C_{\ell}$, see Eq.~(\ref{eq:clfull}), in two different ways: suppressing galaxy clustering and the growth of structure via $P_m(k)$ as well as modifying the background expansion rate via the comoving distance which appears in the argument of the Bessel function $j_\ell$. Among these two effects (i.e. growth versus background) we find that the growth suppression effects in the matter power spectrum due to the presence of massive neutrinos will dominate over background effects. Therefore, the neutrino mass constraints presented in the following analysis arise mostly from the suppression of clustering rather than from purely geometrical effects.     

\section{Mocks}
\label{sec:mocks}

We first consider angular spectra based on mock galaxy catalogs to test 
that neither our method of estimating the spectra nor our modeling 
of the spectra introduces a bias in the reconstructed cosmological parameters. 
We use twenty
independent CMASS mock catalogs based on
N-body simulations and a Halo Occupation Distribution (HOD) model 
described in \cite{whiteetal11} (see also our companion papers \cite{hoetal11,seobao} for details). The input cosmology
for the simulations is $\Omega_{\rm DM} = 0.274, h = 0.7, \Omega_b = 0.046, n_s = 0.95, \sigma_8 = 0.8$ in a
spatially flat universe,
with $n_s$ and $\sigma_8$ the scalar spectral index and the linear rms density fluctuations in
spheres of radius $8 \, h^{-1}$Mpc at $z=0$, respectively. Neutrinos are massless in the input cosmology. The catalogs cover a 
cubic volume with side $1.5 h^{-1}$~Gpc.
To construct an ``observed'' catalog, we put the observer in one corner of the box and consider
the subsample of galaxies in the shell octant between the observer's $z = 0.5 - 0.55$. For simplicity,
we do not apply photometric redshift errors nor do we introduce redshift shifts due to peculiar
velocities. This latter effect would only be significant on very large scales anyway (see Figure \ref{fig: data}).
Each mock covers $\pi/2$ rad$^2$ and consists of about 125,000 galaxies. Since both area and galaxy number are thus roughly half
the values for the $z=0.5-0.55$ redshift bin of the true data, the number density
is comparable to that of the true photometric sample.
We apply this procedure for eight different corners to get eight different lines of
sight per simulation. Note, however, that these lines of sight are not completely independent as
they are based on the same simulation volume.
Finally, for each line of sight, we average the spectra over all twenty independent realizations
in order to increase the signal to noise ratio. The covariance matrix for the mock angular power spectrum
is rescaled accordingly to reflect the decrease in covariance due to taking the average.

\begin{figure*}[!htb]
\centering
  \includegraphics*[width=\linewidth]{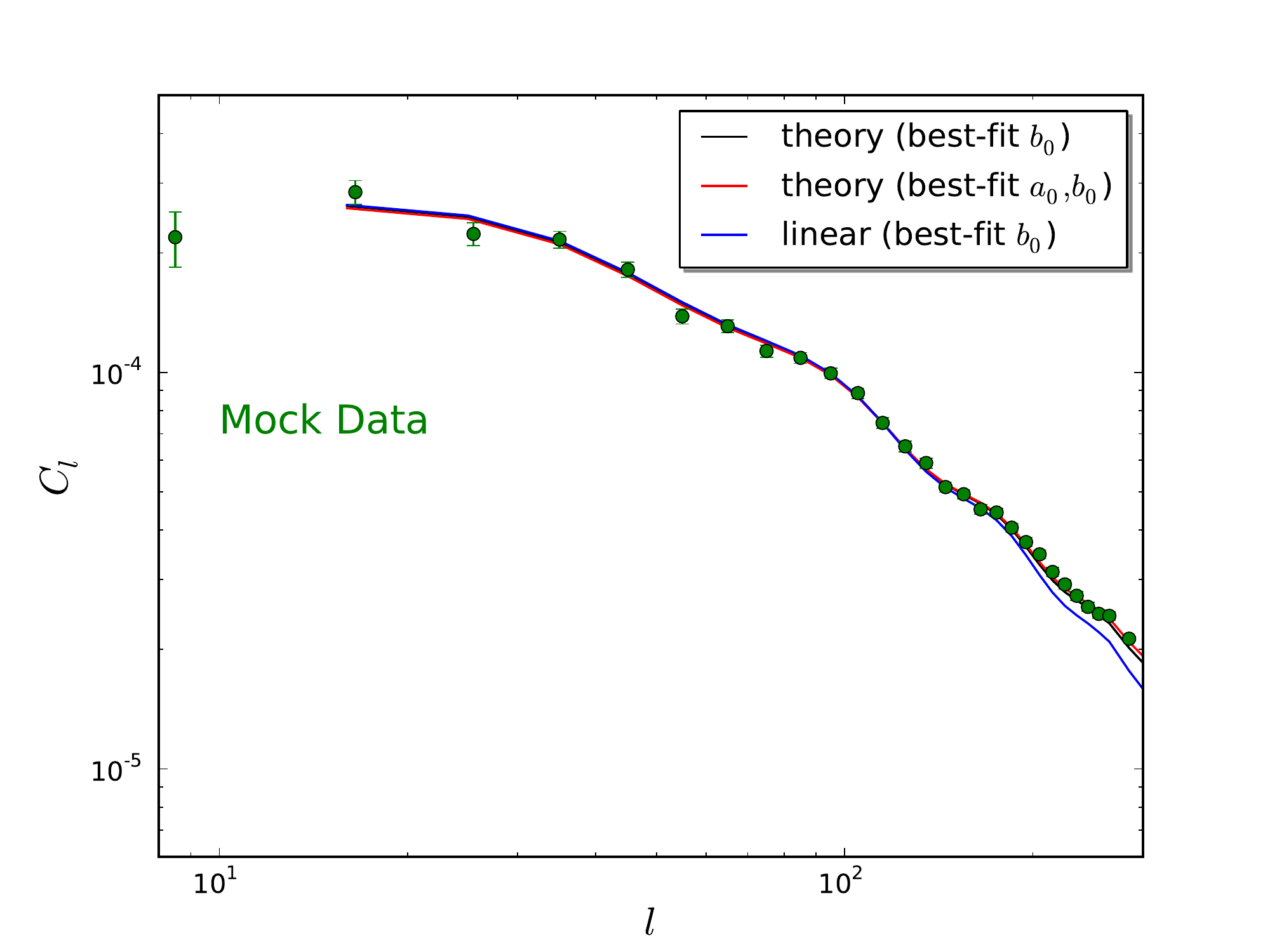}
  \caption{Example of an averaged mock spectrum (green points with error bars) and theoretical spectra (solid lines).
Fixing the cosmology to the mock input cosmology (see text), we fit the averaged mock spectrum in the range
$\ell = 30-200$ to our model described in the text. The black curve is the resulting best-fit spectrum if
we only fit a (scale-independent) galaxy bias $b_0$ (best-fit value $b_0 = 2.02$), while the red curve is the best fit
in a model that also includes the nuisance parameter $a_0$ (best fit values $b_0=2.00, \, a_0=1.05 \cdot 10^{-6}$). To provide
a hint of the importance of non-linear effects in this multipole range, we plot the spectrum based on a linear three
dimensional matter power spectrum in blue ($b_0 = 2.02, a_0=0$)
}
  \label{fig: mock spec}
\end{figure*}

As described in section \ref{sec:model}, we consider a model characterized by
the cosmological parameters and a galaxy bias $b_0$ (giving our mock bin the label $0$),
and a more conservative model with bias $b_0$ {\it and} nuisance parameter $a_0$,
so that the spectrum is given by
\beq
\label{eq:clfull3}
C_\ell^{(00)} = b_0^2 \, \frac{2}{\pi} \int k^2 dk \, P_m(k,z=0) \, \left( \Delta_\ell^{(i)}(k) \right)^2 + a_0~.
\eeq
In the galaxy bias-only version, $a_0$ is simply set to zero.

As a direct test of this model,
we fit it to the averaged mock spectrum.
In this first approach, we keep the cosmology fixed to the mock's input cosmology and restrict
the fit to the range $\ell=30-200$. The only free parameters are thus either $b_0$
or $(b_0, a_0)$. We use a modification of the publicly available COSMOMC package \cite{LewisBridle02} to sample this parameter
space using Monte Carlo Markov Chains (MCMC).
We show the resulting best-fit spectra together with the mock average in
Fig.~\ref{fig: mock spec}. Considering first the default, galaxy bias-only model (black curve),
we find that the best fit to the mock result has a linear bias $b_0 = 2.02$ (with uncertainty
$\sigma(b_0) < 0.01$) and has $\chi^2 = 11.3$. This should be compared to an expectation value of
$\langle \chi^2 \rangle = 16$ based on 17 data bins and one free parameter. The galaxy bias model thus provides
a good fit to the simulated spectrum (the probability of getting a $\chi^2$ lower than 11.3
for an expectation value of 16 is approximately $20 \%$).

Next, including the shot noise-like parameter $a_0$ to take into account potential residual shot noise and/or
non-linear effects not captured
by our simple Halofit plus scale-independent galaxy bias model, we find a best fit model with $a_0=1.1 \cdot 10^{-6}$ and
$b_0 = 2.00$. However, the uncertainty in $a_0$ is $\sigma(a_0) = 1.0 \cdot 10^{-6}$ so the preference for a non-zero
value cannot be considered significant. In this model, we find $\chi^2 = 10.1$, to be compared
to the expected $\langle \chi^2 \rangle = 15$.
This is only a marginal ($\Delta \chi^2 \approx 1.2$) improvement.

Restricting the fitting range to $\ell = 30 - 150$, we find $\Delta \chi^2 = 1.1$ between the two best-fit models,
and $a_0 = (1.8 \pm 1.9) \times 10^{-6}$.

\begin{figure*}[!htb]
\begin{minipage}[t]{0.49\textwidth}
\centering
  \includegraphics*[width=\linewidth]{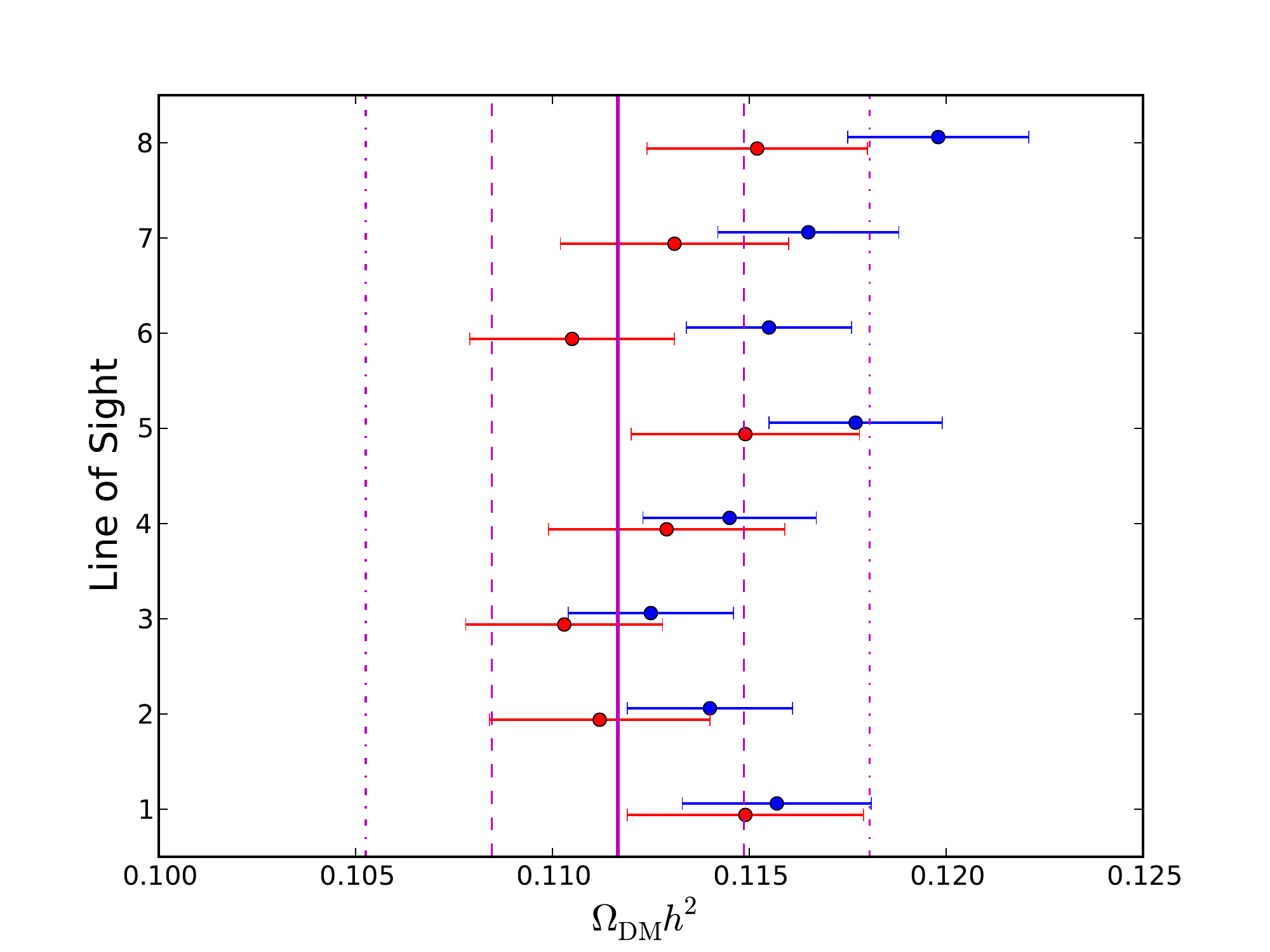}
\end{minipage} \hfill
\begin{minipage}[t]{0.49\textwidth}
\centering
  \includegraphics*[width=\linewidth]{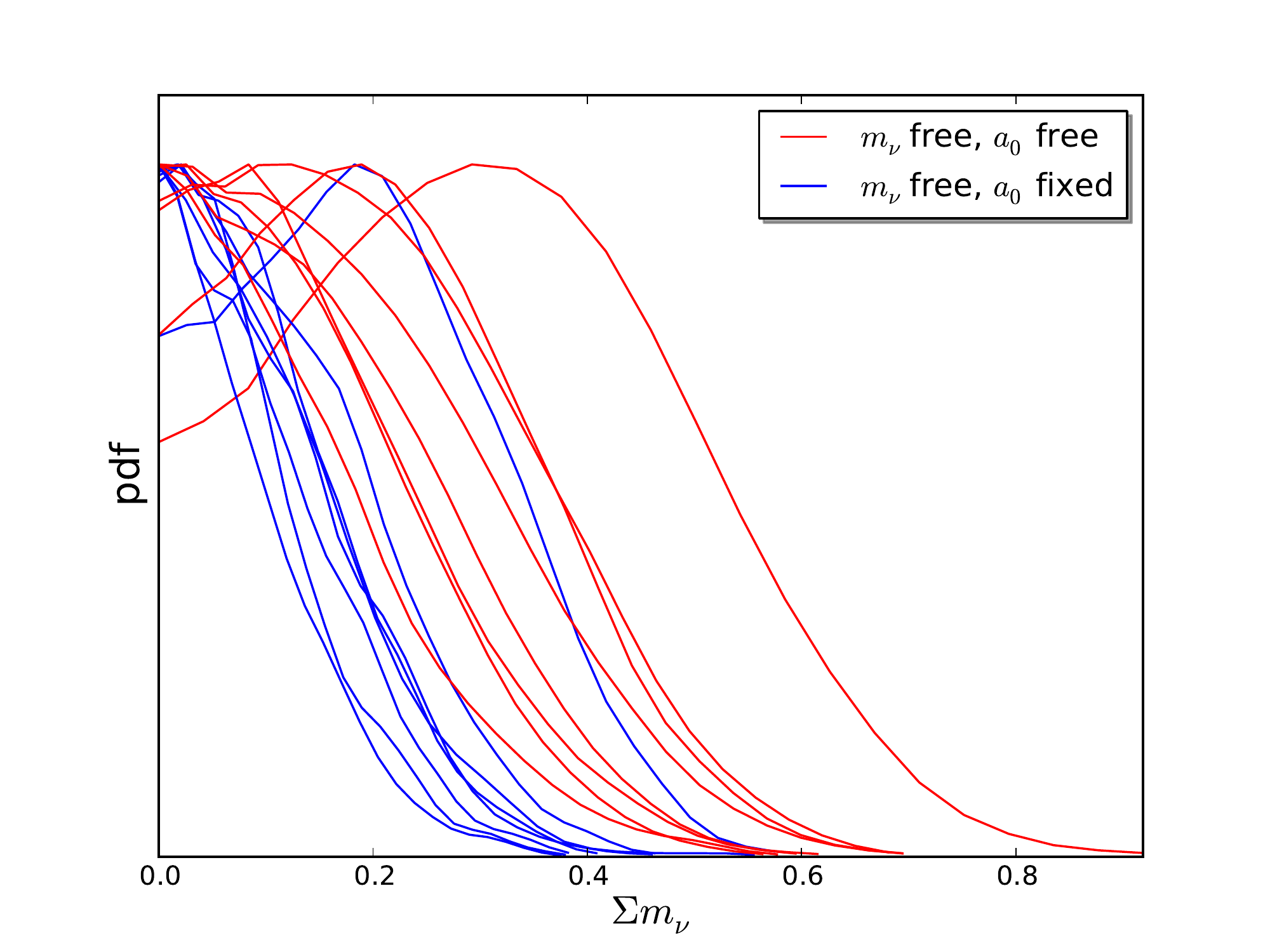}
\end{minipage}
  \caption{{\em Left panel:} 
  Recovered values of $\Omega_{\rm DM} h^2$ from
  averaged mock spectrum together with CMB prior. We consider spectra from eight different
  lines of sight.
  The points with error bars show the
  posterior mean values and $1\sigma$ error bars
  after the Monte Carlo analysis for two scenarios: varying $\Sigma m_{\nu}$
  {\it with} (red) and {\it without} (blue) $a_0$ marginalized. The vertical magenta lines indicate the input
  value $\Omega_{\rm DM}h^2=0.11166$ (solid) and the input $\pm$ one and two $\sigma$,
  where $\sigma=0.0032$ is the parameter uncertainty
  based on the data set WMAP7+HST+angular spectra ($a_i=0$ fixed).
  There is a bias of about $1\sigma$ without the nuisance parameter,
  which disappears when $a_i$ is marginalized over.
  {\em Right panel:} The posterior neutrino mass distributions for the two cases discussed
  above. The mock constraints are consistent with the input cosmology
  of $\Sigma m_{\nu} = 0$.
  Other parameters are all reconstructed to close to their input values and are not strongly affected
  by the angular spectra.
}
  \label{fig:mockcosmo}
\end{figure*}

The comparison above suggests that for the range $\ell=30-200$, the galaxy bias model without an extra nuisance
parameter may be sufficient. We now undertake a more complete check of our model and the entire cosmology analysis
by using MCMC
to fit the full space of cosmological parameters as well as the galaxy bias (and shot noise parameter) to the
averaged mock spectrum. The differences between the resulting best-fit parameter values
and the ``true'', i.e.~input, values give an indication of the parameter bias introduced by our method.
To break parameter degeneracies, while not letting the prior bias us away
from the input cosmology, we include a ``mock''
CMB prior\footnote{ The ``mock'' CMB prior is
defined by $\chi^2_{\rm WMAP7} \equiv (p_i - p_{i, {\rm input}})\,{\rm COV}_{ij}^{-1}\,(p_j - p_{, {\rm input}})$,
where $p_i$ are the parameters at each point of the chain, $p_{i, {\rm input}}$ the input parameters, ${\rm COV}_{ij}$
is the WMAP7 covariance matrix and $i, j$ are summed over.}, which will provide a likelihood similar to the
true WMAP7 one, except shifted to be centered around the mock input parameters.

We want any deviation between the input cosmology and the recovered cosmology to be small
compared to the uncertainty level of the actual data.
We therefore take information from our final results and
use the uncertainties for the WMAP+HST+CMASS ($\ell_{\rm max}=200$)
real data case for comparison. If the biases on parameter estimation
are small compared
to these numbers, it provides strong motivation
for considering our approach sound, as most uncertainties in the next section will be larger than in
the WMAP+HST+CMASS case. Therefore, parameter uncertainties $\sigma$ referred to in the remainder of this section
are these data-based uncertainties.

Starting with the parameter space\footnote{The parameters $\theta$, $A_s$ and $\tau$
represent the ratio between the sound horizon and the angular diameter distance at decoupling,
the scalar amplitude of primordial fluctuations and the reionization optical depth, respectively.}
$\{\Omega_b h^2, \Omega_{\rm DM} h^2, \theta, A_s, n_s, \tau, b_0 \}$,
we find that all cosmological parameters are reproduced to within
$1\sigma$ of the input values (although the parameter most affected by the mock CMASS data, $\Omega_{\rm DM} h^2$,
is higher than the input by close to $1\sigma$).

Unfortunately, we do not have mocks based on a cosmology with non-zero $\Sigma m_{\nu}$.
One check we {\it can} do, however, is to fit a model with parameters
$\{\Omega_b h^2, \Omega_{\rm DM} h^2, \theta, A_s, n_s, \tau, \Sigma m_{\nu}, b_0 \}$
to our $\Sigma m_{\nu} = 0$ mock spectra. The parameters affected by far the most by the angular spectra
are (again) $\Omega_{\rm DM} h^2$ and $\Sigma m_{\nu}$.
We show the posteriors of this calculation in Fig.~\ref{fig:mockcosmo}.
In the left panel, the vertical lines indicate the $\Omega_{\rm DM} h^2$ input value, and the
$1\sigma$ and $2\sigma$ bounds based on the uncertainty $\sigma$ from the actual data.
The blue points with error bars are the posterior mean values and $1\sigma$ recovered errors after fitting to the averaged
mock spectrum.
Note that the recovered error bars (from the averaged mock power spectrum)
are typically similar to the data-based
error bars.
While the different lines of sight are not entirely independent,
Fig.~\ref{fig:mockcosmo} points towards a bias of about $1 - 1.5 \sigma$ in $\Omega_{\rm DM}h^2$.
For the neutrino mass, the right panel shows the posterior probability distributions in blue.
The posteriors are always consistent with the input value $\Sigma m_{\nu}=0$
and can be interpreted as providing upper bounds.
We have made the same plot as in the left panel for the other parameters and they were biased significantly less
(as their reconstruction is dominated by the mock CMB prior).

Adding the nuisance parameter $a_0$, we obtain the red points and curves in Fig.~\ref{fig:mockcosmo}.
The effect of marginalizing over $a_0$ is to diminish the parameter bias so
that $\Omega_{\rm DM} h^2$ is typically reconstructed to well within $1\sigma$.
We attribute this change to $a_0$ accounting for
a possible scale-dependence in galaxy bias on quasilinear scales.
The neutrino constraints are also still consistent with the input, although the
mock upper limits do become significantly weaker.
We have also studied mock cosmology constraints using $\ell_{\rm max}=150$,
and found that the main effect is to widen the posterior distributions slightly,
while the change in parameter bias relative to $\ell_{\rm max}=200$
is small.

We conclude that our galaxy bias-only model and the fitting method used here properly
reproduce the input cosmology for our choices of $\ell_{\rm max}$,
except that there is a bias of about $1 - 1.5 \sigma$ in $\Omega_{\rm DM} h^2$.
The model with nuisance parameter $a_0$ removes this parameter bias at the cost of larger error bars.
While the bias in $\Omega_{\rm DM}h^2$ is not extreme, being only slightly above the $1\sigma$ level,
it is sufficiently worrying that we will quote results for the galaxy bias-only model
{\it and} for the more conservative model with shot noise-like parameters.
Changing $\ell_{\rm max}$ between 150 and 200 does not have a large effect on how well
the models compare to mocks, suggesting that both are reasonable choices. We will quote results for
both ranges.

\section{Results}
\label{sec:results}

While the CMASS angular galaxy power spectra carry useful information about the sum of neutrino
masses $\Sigma m_{\nu}$, the effect of $\Sigma m_{\nu}$ is degenerate with certain other parameters which are not
well constrained from the angular spectra alone.
There are many combinations of external data sets that our angular spectra can be combined with
to fix this problem.
One approach would be to optimize the neutrino bound by combining as many data sets as possible.
However, we choose instead to focus as much as possible on the effect of the CMASS photometric data
and therefore consider mostly simple priors. Our two main prior choices are WMAP7 CMB data~\cite{Larsonetal10} and
the combination of WMAP7 with the HST measurement of the Hubble parameter~\cite{hst11}.
At the end of this section, we will briefly consider the effect of adding the Union 2
supernova compilation \cite{Union2} and the measurement of the BAO scale
based on SDSS Data Release 7 \cite{DR7} spectroscopic data from Ref.~\cite{Percetal10}. 

We again use a modification of the publicly available COSMOMC package \cite{LewisBridle02} to sample the parameter
space using MCMC. Our parameter space consists of the six usual
$\Lambda$CDM parameters, $(\Omega_b h^2, \Omega_{\rm DM} h^2, \theta, \ln(10^{10}\,A_s), n_s, \tau)$, the neutrino mass fraction
$f_\nu$, defined as $\Omega_{\nu}/\Omega_{\rm DM}$ (where $\Omega_{\rm DM}$ includes cold dark matter and massive neutrinos),
in addition to $A_{\rm SZ}$, describing the amplitude relative to a template of the Sunyaev-Zel'dovich contribution to the CMB~\cite{komsel02},
the four galaxy bias parameters $b_i$ and (optionally) the four nuisance
parameters $a_i$, leaving us with a maximum total number of parameters of sixteen parameters. We put uniform priors on these parameters
and derive $\Sigma m_{\nu}$ using Eq.~(\ref{eq:omnu}).

\begin{table*}[hbt!]
\begin{center}
\small
\begin{tabular}{c|ccc}
\hline\hline
$95 \%$ CL $\sum m_{\nu} $[eV$]$ & prior only & prior+CMASS,$\ell_{\rm max}=150$ &  prior+CMASS,$\ell_{\rm max}=200$\\
\hline\hline
WMAP7 prior & 1.1 & 0.74 (0.92) & 0.56 (0.90)\\
\hline
WMAP7 + HST prior & 0.44 &  0.31 (0.40) & 0.26 (0.36) \\
\hline\hline
\end{tabular}
\caption{The $95 \%$ confidence level upper limits on the sum of the neutrino
masses $\Sigma m_{\nu}$. The top row investigates the effect of adding the CMASS galaxy power spectra
to a WMAP prior while the bottom row uses WMAP and the $H_0$ constraint from HST as a prior.
In parentheses we show results for the more conservative model marginalizing over the shot noise-like
parameters $a_i$.
}
\label{tab:results}
\end{center}
\end{table*}

\begin{figure*}[!htb]
\centering
  \includegraphics*[width=\linewidth]{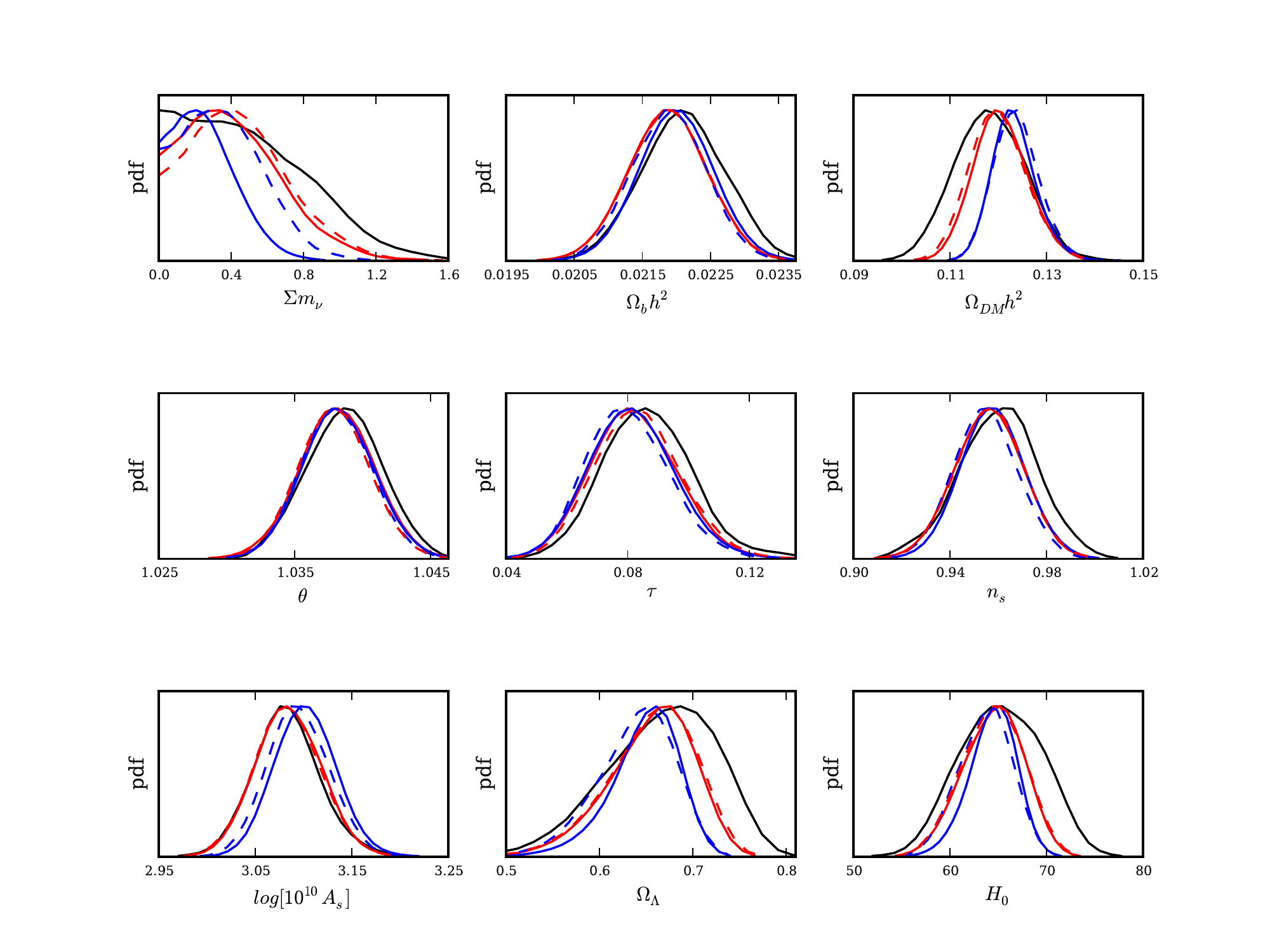}
  \caption{Cosmological constraints with a WMAP7 CMB prior. We show the probability distribution
  functions for CMB only (black), CMB with CMASS spectra in the range $\ell=30-150$ (blue dashed)
  and CMB with CMASS spectra in the range $\ell=30-200$ (blue solid).
  The red curves represent the constraints in the conservative model where we marginalize
  over a set of nuisance parameters $a_i$.
}
  \label{fig:1dresultsnohst}
\end{figure*}

\begin{figure*}[!htb]
\centering
  \includegraphics*[width=\linewidth]{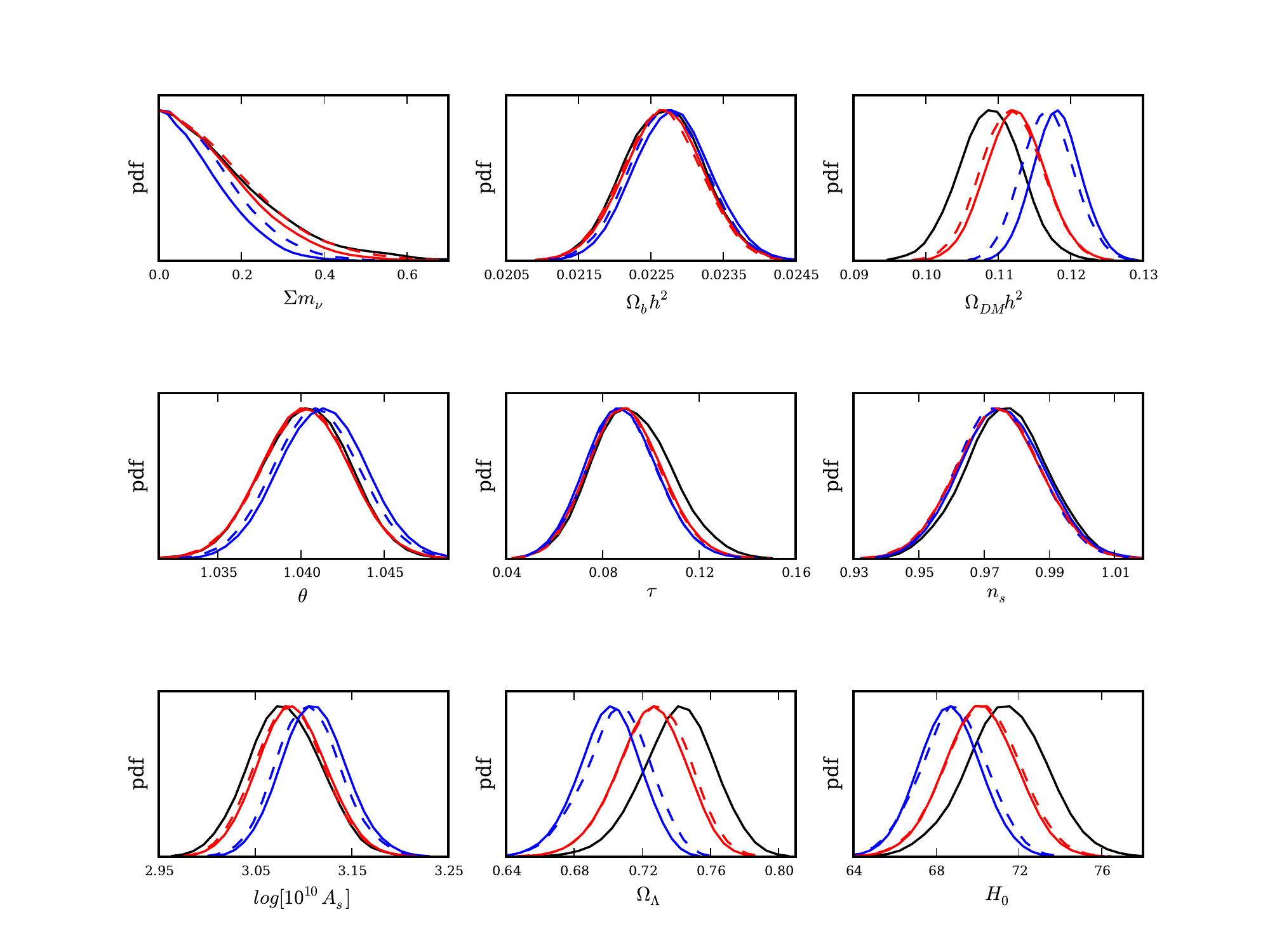}
  \caption{Cosmological constraints with a WMAP7 CMB {\it and} Hubble parameter prior. We show the probability distribution
  functions for CMB + $H_0$ only (black), CMB + $H_0$ with CMASS spectra in the range $\ell=30-150$ (blue dashed)
  and CMB + $H_0$ with CMASS spectra in the range $\ell=30-200$ (blue solid).
  The red curves represent the constraints in the conservative model where we marginalize
  over a set of nuisance parameters $a_i$. While this marginalization degrades the neutrino bound,
  simulations have shown it removes the bias in $\Omega_{\rm DM}h^2$ (see section \ref{sec:mocks}).
}
  \label{fig:1dresultshst}
\end{figure*}

We first consider the WMAP7 prior and show how the neutrino bound improves as CMASS data are added.
The resulting $95 \%$ CL upper limits are shown in the top row of Table \ref{tab:results},
with the results with $a_i$ marginalized in parentheses. The bound
improves from $1.1$ eV for CMB only to $0.56$ eV for CMB with CMASS data ($\ell_{\rm max}=200$).
This constraint is comparable to the limit $\Sigma m_{\nu} < 0.62$ eV derived by \cite{Reid:2009xm} from the DR7 spectroscopic sample.
It thus appears that the advantage of spectroscopic redshifts (providing information on clustering along the line of sight)
in that sample is offset by the advantage of the current sample having a larger volume,
although there are other differences between the samples and analyses as well.
Note, however, that the constraint deteriorates significantly when marginalizing over the nuisance parameters $a_i$.
In this case, the mass bound is not significantly better than with CMB alone.
We show the posterior probability distributions for $\Sigma m_{\nu}$ and the other cosmological parameters
in Fig.~\ref{fig:1dresultsnohst}.

We next consider the constraints using WMAP7 with HST $H_0$ prior. The CMB alone provides a strong measurement of
one combination of late-universe parameters through its sensitivity to the distance to the last scattering
surface. However, this distance measurement leaves a degeneracy between $\Omega_{\Lambda}$ and $\Sigma m_{\nu}$
so that the CMB-only limit on the neutrino mass arises mainly from the effect of neutrinos on the primary anisotropies
and not from this distance measurement. Measuring $H_0$ constrains a different combination of late universe parameters
and thus breaks the CMB degeneracy. This is why the WMAP7+HST bound is so much stronger than the WMAP7-only one,
i.e.~$\Sigma m_{\nu} < 0.44$~eV as opposed to $\Sigma m_{\nu} < 1.1$~eV.
Adding the CMASS angular spectra tightens the bound significantly
so that an impressive upper bound of $\Sigma m_{\nu} < 0.26$~eV is reached for $\ell_{\rm max}=200$
(in the bias-only model), as is shown in the second row of Table \ref{tab:results}.
The effect of marginalization over $a_i$ is again to bring the constraint back to closer
to the CMB+HST bound.

The posteriors for all cosmological parameters are shown in Fig.~\ref{fig:1dresultshst}.
In addition to the full likelihoods for $\Sigma m_{\nu}$, summarized in Table \ref{tab:results},
the $\Omega_{\rm DM}h^2$ posteriors are worth noting. The effect of the angular spectra is to strongly
shift the average value of this parameter (blue curve), while including the nuisance parameters (red curve) weakens
the shift.
The last two parameters in Fig.~\ref{fig:1dresultshst} (and \ref{fig:1dresultsnohst}),
$\Omega_\Lambda$ and $H_0$,
are not independent and can be expressed in terms of the preceding parameters.
The shift thus is really only significant for one independent parameter, $\Omega_{\rm DM}h^2$,
in our basis.
The results in section \ref{sec:mocks} suggest that the shift in the bias-only case
might partially be a bias due to our model
and that the results with $a_i$ marginalized are unbiased.

We do not explicitly show the correlations
between parameters, but have verified that, in the CMB+CMASS case,
the neutrino mass has its strongest degeneracies with $\Omega_{\rm DM}h^2$,
the bias parameters $b_i$ and $\sigma_8$. While,
in agreement with our discussion in section \ref{sec:nu},
the inclusion of the Hubble prior removes the $\Sigma m_{\nu} - \Omega_{\rm DM}h^2$
degeneracy, the strong correlations with $b_i$ and $\sigma_8$ remain.

We have also added
supernova and BAO data to the CMB+HST+CMASS data set, and considered the neutrino mass bound
in the bias-only model, but we found negligible
improvement (from $0.26$~eV to $0.25$~eV) relative to the case without
these additional data sets. These additional data sets do carry significant information,
but this information is degenerate with the information already present
in the three default data sets.

For the multipole range $\ell = 30 - 150$, we show the results using dashed lines in Figures \ref{fig:1dresultsnohst} and \ref{fig:1dresultshst}.
The $95 \%$ CL
upper limit for CMB+HST+CMASS in this case is $0.31 (0.40)$ eV and for CMB+CMASS it is $0.74 (0.92)$ eV fixing (varying) $a_i=0$.
A significant amount of information is thus contained in the large $\ell$ range of multipoles, which makes sense
as the number of modes is large.

Finally, we consider the question of where most of the neutrino mass information comes from.
In principle, massive neutrinos affect the angular power spectra both by their small-scale
suppression of the three-dimensional power spectrum, and by changing the projection of physical scales
onto angular scales through their effect on the background expansion. As discussed in section \ref{sec:nu},
we expect the former effect to carry more constraining power than the latter effect. We test this by
running Monte Carlo chains where the effect of massive neutrinos on the three-dimensional power spectrum
is artificially taken out, while
the effect on the background expansion is left intact.
Specifically, we replace
the usual linear CAMB power spectrum by the spectrum
given by the Eisenstein and Hu (EH) fitting formula \cite{eishu98}, which does not
include the effect of massive neutrinos. We find that in this setup, including the
CMASS galaxy power spectra does {\it not} improve the neutrino mass bound relative to the case
with CMB, or with CMB+HST, only. In other words, the projection effect alone carries little information
on neutrino mass (at least after marginalizing over the effects of other parameters)
and the bounds quoted in this manuscript can be attributed to the small-scale suppression of the three-dimensional
power spectrum.

\section{Conclusions}
\label{sec:disc}

We have exploited angular power spectra from the SDSS-III DR8 sample
of photometric galaxies with CMASS selection criteria to put interesting constraints on the sum
of neutrino masses.
We have used mock galaxy catalogs based on N-body simulations and HOD modeling
to test two models for the angular galaxy spectra.
Based on these tests, we decided to compare the data
to theoretical spectra based on the non-linear
matter power spectrum augmented by a linear galaxy bias factor.
However, since this model does result in a bias in $\Omega_{\rm DM} h^2$
of $\sim 1 - 1.5 \sigma$, we have also fitted the data to a more conservative
model, with an additional set of shot noise-like fitting parameters,
in which this bias is virtually absent.
The tests also motivated us to use the multipole
range $\ell=30-200$, but we quoted results
for the more conservative choice $\ell=30-150$ as well.
The added advantage is that this analysis provides insight into the range of scales
that yields the galaxy clustering information.

Combining the CMASS data with a CMB prior from the WMAP7 survey,
we find an upper bound $\Sigma m_{\nu} < 0.56$ eV ($0.90$ eV) at $95 \%$
confidence level for $\ell_{\rm max} = 200$ in the model with free parameters
$b_i$ ($b_i$ and $a_i$). Adding the HST
measurement of the Hubble parameter, the probability distribution tightens
and we find $\Sigma m_{\nu} < 0.26$ eV ($0.36$ eV).
We have also considered the effect of adding supernova and a (lower redshift)
BAO measurement, but when the HST prior is included already, these additions lower the upper limit
to $0.25$ eV (in the bias-only model).
Considering the dependence on the multipole range, characterized by a maximum multipole $\ell_{\rm max}$,
we find that a significant amount of information resides in the largest multipoles $\ell=150-200$,
but that even for $\ell_{\rm max}=150$, the galaxy spectra place a strong bound on neutrino mass.
Our main results are summarized in Table \ref{tab:results}.

It is interesting to compare these results to the outcome of an analysis
of a similar (but smaller) high redshift SDSS photometric catalog, the MegaZ
sample \cite{megaz}. In \cite{Thomas:2009ae}, the strongest bound quoted is a $95 \%$ CL
upper limit of $0.28$ eV, including SN and BAO data in addition to
CMB, HST and MegaZ. However, this particular bound is based on a multipole range with
$\ell_{\rm max} = 300$ and no nuisance parameters $a_i$. As we have discussed extensively,
we believe $\ell_{\rm max} = 200$ (or even slightly lower) is a better choice
if one wants to avoid significant, unknown non-linear corrections to the galaxy bias.
For this choice, the MegaZ sample yields an upper bound of $0.34$ eV.
Assuming the aggressive, bias-only model (as in the MegaZ analysis).
the value we find for the CMASS sample is $0.25$ eV, which is thus a significant
improvement. However, it must be kept in mind that this model causes a small
parameter bias and that the more conservative model yields a weaker bound
of $0.36$ eV.

The bounds presented here rule out
the quasi-degenerate neutrino mass hierarchy. For example, for $\Sigma m_{\nu} = 0.25$ eV, it follows
from $|\Delta m^2_{23}|=2.5\cdot 10^{-3}$~eV$^2$ that the mass difference $|m_3 - m_2| \approx 0.015$ eV,
so that the largest mass difference is $|m_3-m_2|/(\Sigma m_{\nu}/3) \approx 20 \%$ of
the average neutrino mass. We are thus entering the regime where the mass splittings are significant.
Looking forward, the prospects are exciting.
As the sensitivity of cosmological neutrino mass measurements improves,
the sum of the masses will either be measured, i.e.~a value that can be distinguished from zero will be found,
or the upper limit will be sharpened. However, even in the latter case, interesting things can be learned. If the sum of the masses
is found to be less than $\sim 0.1$eV, this rules out the inverted hierarchy, leaving the normal hierarchy as the only option.
Moreover, even in the normal hierarchy, $\Sigma m_{\nu}$ is not allowed to be lower than $\sim 0.05$eV
so that sooner or later a measurement, rather than an upper bound, can be expected.

Finally, we note that BOSS is currently taking spectra
for a sample of galaxies with the same selection criteria as the galaxies
considered in this paper and will reach a sample of approximately $1,500,000$
galaxies when finished in 2014. The results presented here and in our companion papers
are thus only the tip of the iceberg of what can be done with BOSS.
The spectroscopic data will allow a measurement of the three-dimensional
power spectrum for an even larger volume than considered here (if the low redshifts
sample is included) so that also clustering in the line-of-sight direction can be resolved,
thus promising significantly stronger cosmology constraints than from the photometric
data.

\bigskip
\noindent{\bf Acknowledgement}:

Funding for SDSS-III has been provided by the Alfred P. Sloan Foundation, the Participating Institutions, the National Science Foundation, and the U.S. Department of Energy Office of Science. The SDSS-III web site is http://www.sdss3.org/.

SDSS-III is managed by the Astrophysical Research Consortium for the Participating Institutions of the SDSS-III Collaboration including the University of Arizona, the Brazilian Participation Group, Brookhaven National Laboratory, University of Cambridge, University of Florida, the French Participation Group, the German Participation Group, the Instituto de Astrofisica de Canarias, the Michigan State/Notre Dame/JINA Participation Group, Johns Hopkins University, Lawrence Berkeley National Laboratory, Max Planck Institute for Astrophysics, New Mexico State University, New York University, Ohio State University, Pennsylvania State University, University of Portsmouth, Princeton University, the Spanish Participation Group, University of Tokyo, University of Utah, Vanderbilt University, University of Virginia, University of Washington, and Yale University.

RdP is supported by FP7-IDEAS-Phys.LSS 240117.
O.M. is supported by AYA2008-03531 and the Consolider Ingenio project CSD2007-00060.


\bibliographystyle{apj}
\bibliography{refs}

\begin{thebibliography}{78}
\expandafter\ifx\csname natexlab\endcsname\relax\def\natexlab#1{#1}\fi

\bibitem[{{Abazajian} {et~al.}(2009){Abazajian}, {Adelman-McCarthy},
  {Ag{\"u}eros}, {Allam}, {Allende Prieto}, {An}, {Anderson}, {Anderson},
  {Annis}, {Bahcall}, \& et~al.}]{DR7}
{Abazajian}, K.~N., {Adelman-McCarthy}, J.~K., {Ag{\"u}eros}, M.~A., {et~al.}
  2009, \apjs, 182, 543

\bibitem[{{Aihara} {et~al.}(2011){Aihara}, {Allende Prieto}, {An}, {Anderson},
  {Aubourg}, {Balbinot}, {Beers}, {Berlind}, {Bickerton}, {Bizyaev}, {Blanton},
  {Bochanski}, {Bolton}, {Bovy}, {Brandt}, {Brinkmann}, {Brown}, {Brownstein},
  {Busca}, {Campbell}, {Carr}, {Chen}, {Chiappini}, {Comparat}, {Connolly},
  {Cortes}, {Croft}, {Cuesta}, {da Costa}, {Davenport}, {Dawson}, {Dhital},
  {Ealet}, {Ebelke}, {Edmondson}, {Eisenstein}, {Escoffier}, {Esposito},
  {Evans}, {Fan}, {Femen{\'{\i}}a Castell{\'a}}, {Font-Ribera}, {Frinchaboy},
  {Ge}, {Gillespie}, {Gilmore}, {Gonz{\'a}lez Hern{\'a}ndez}, {Gott}, {Gould},
  {Grebel}, {Gunn}, {Hamilton}, {Harding}, {Harris}, {Hawley}, {Hearty}, {Ho},
  {Hogg}, {Holtzman}, {Honscheid}, {Inada}, {Ivans}, {Jiang}, {Johnson},
  {Jordan}, {Jordan}, {Kazin}, {Kirkby}, {Klaene}, {Knapp}, {Kneib},
  {Kochanek}, {Koesterke}, {Kollmeier}, {Kron}, {Lampeitl}, {Lang}, {Le Goff},
  {Lee}, {Lin}, {Long}, {Loomis}, {Lucatello}, {Lundgren}, {Lupton}, {Ma},
  {MacDonald}, {Mahadevan}, {Maia}, {Makler}, {Malanushenko}, {Malanushenko},
  {Mandelbaum}, {Maraston}, {Margala}, {Masters}, {McBride}, {McGehee},
  {McGreer}, {M{\'e}nard}, {Miralda-Escud{\'e}}, {Morrison}, {Mullally},
  {Muna}, {Munn}, {Murayama}, {Myers}, {Naugle}, {Fausti Neto}, {Cuong Nguyen},
  {Nichol}, {O'Connell}, {Ogando}, {Olmstead}, {Oravetz}, {Padmanabhan},
  {Palanque-Delabrouille}, {Pan}, {Pandey}, {P{\^a}ris}, {Percival},
  {Petitjean}, {Pfaffenberger}, {Pforr}, {Phleps}, {Pichon}, {Pieri}, {Prada},
  {Price-Whelan}, {Raddick}, {Ramos}, {Reyl{\'e}}, {Rich}, {Richards}, {Rix},
  {Robin}, {Rocha-Pinto}, {Rockosi}, {Roe}, {Rollinde}, {Ross}, {Ross},
  {Rossetto}, {S{\'a}nchez}, {Sayres}, {Schlegel}, {Schlesinger}, {Schmidt},
  {Schneider}, {Sheldon}, {Shu}, {Simmerer}, {Simmons}, {Sivarani}, {Snedden},
  {Sobeck}, {Steinmetz}, {Strauss}, {Szalay}, {Tanaka}, {Thakar}, {Thomas},
  {Tinker}, {Tofflemire}, {Tojeiro}, {Tremonti}, {Vandenberg}, {Vargas
  Maga{\~n}a}, {Verde}, {Vogt}, {Wake}, {Wang}, {Weaver}, {Weinberg}, {White},
  {White}, {Yanny}, {Yasuda}, {Yeche}, \& {Zehavi}}]{DR8}
{Aihara}, H., {Allende Prieto}, C., {An}, D., {et~al.} 2011, \apjs, 193, 29

\bibitem[{Allen {et~al.}(2003)Allen, Schmidt, \& Bridle}]{Allen:2003pta}
Allen, S., Schmidt, R., \& Bridle, S. 2003, Mon.Not.Roy.Astron.Soc., 346, 593

\bibitem[{Amanullah~{\it et al}(2010)}]{Union2}
Amanullah~{\it et al}, R. 2010, Astrophys.J. accepted

\bibitem[{Barger {et~al.}(2004)Barger, Marfatia, \& Tregre}]{Barger:2003vs}
Barger, V., Marfatia, D., \& Tregre, A. 2004, Phys.Lett., B595, 55

\bibitem[{{Benson} {et~al.}(2011){Benson}, {de Haan}, {Dudley}, {Reichardt},
  {Aird}, {Andersson}, {Armstrong}, {Bautz}, {Bayliss}, {Bazin}, {Bleem},
  {Brodwin}, {Carlstrom}, {Chang}, {Cho}, {Clocchiatti}, {Crawford}, {Crites},
  {Desai}, {Dobbs}, {Foley}, {Forman}, {George}, {Gladders}, {Halverson},
  {High}, {Holder}, {Holzapfel}, {Hoover}, {Hrubes}, {Jones}, {Joy}, {Keisler},
  {Knox}, {Lee}, {Leitch}, {Liu}, {Lueker}, {Luong-Van}, {Mantz}, {Marrone},
  {McDonald}, {McMahon}, {Mehl}, {Meyer}, {Mocanu}, {Mohr}, {Montroy},
  {Murray}, {Natoli}, {Padin}, {Plagge}, {Pryke}, {Rest}, {Ruel}, {Ruhl},
  {Saliwanchik}, {Saro}, {Schaffer}, {Shaw}, {Shirokoff}, {Song}, {Spieler},
  {Stalder}, {Staniszewski}, {Stark}, {Story}, {Stubbs}, {Suhada}, {van
  Engelen}, {Vanderlinde}, {Vieira}, {Vikhlinin}, {Williamson}, {Zahn}, \&
  {Zenteno}}]{bensonetal11}
{Benson}, B.~A., {de Haan}, T., {Dudley}, J.~P., {et~al.} 2011, ArXiv e-prints

\bibitem[{{Bird} {et~al.}(2011){Bird}, {Viel}, \& {Haehnelt}}]{birdetal11}
{Bird}, S., {Viel}, M., \& {Haehnelt}, M.~G. 2011, ArXiv e-prints

\bibitem[{{Coles} {et~al.}(1999){Coles}, {Melott}, \& {Munshi}}]{colesetal99}
{Coles}, P., {Melott}, A.~L., \& {Munshi}, D. 1999, \apjl, 521, L5

\bibitem[{{Collister} {et~al.}(2007){Collister}, {Lahav}, {Blake}, {Cannon},
  {Croom}, {Drinkwater}, {Edge}, {Eisenstein}, {Loveday}, {Nichol}, {Pimbblet},
  {de Propris}, {Roseboom}, {Ross}, {Schneider}, {Shanks}, \& {Wake}}]{megaz}
{Collister}, A., {Lahav}, O., {Blake}, C., {et~al.} 2007, \mnras, 375, 68

\bibitem[{Crotty {et~al.}(2004)Crotty, Lesgourgues, \& Pastor}]{Crotty:2004gm}
Crotty, P., Lesgourgues, J., \& Pastor, S. 2004, Phys.Rev., D69, 123007

\bibitem[{Eisenstein \& Hu(1997)}]{Eisenstein:1997jh}
Eisenstein, D.~J., \& Hu, W. 1997, Astrophys.J., 511, 5

\bibitem[{{Eisenstein} \& {Hu}(1998)}]{eishu98}
{Eisenstein}, D.~J., \& {Hu}, W. 1998, \apj, 496, 605

\bibitem[{{Eisenstein} {et~al.}(2011){Eisenstein}, {Weinberg}, {Agol},
  {Aihara}, {Allende Prieto}, {Anderson}, {Arns}, {Aubourg}, {Bailey},
  {Balbinot}, \& et~al.}]{eisetal11}
{Eisenstein}, D.~J., {Weinberg}, D.~H., {Agol}, E., {et~al.} 2011, \apj, 142,
  72

\bibitem[{Eitel(2005)}]{Eitel:2005hg}
Eitel, K. 2005, Nucl.Phys.Proc.Suppl., 143, 197

\bibitem[{Elgaroy \& Lahav(2005)}]{Elgaroy:2004rc}
Elgaroy, O., \& Lahav, O. 2005, New J.Phys., 7, 61

\bibitem[{Elgaroy {et~al.}(2002)Elgaroy, Lahav, Percival, Peacock, Madgwick,
  {et~al.}}]{Elgaroy:2002bi}
Elgaroy, O., Lahav, O., Percival, W., {et~al.} 2002, Phys.Rev.Lett., 89, 061301

\bibitem[{{Fisher} {et~al.}(1994){Fisher}, {Scharf}, \& {Lahav}}]{fisheretal94}
{Fisher}, K.~B., {Scharf}, C.~A., \& {Lahav}, O. 1994, \mnras, 266, 219

\bibitem[{Fogli {et~al.}(2008)Fogli, Lisi, Marrone, Melchiorri, Palazzo,
  {et~al.}}]{Fogli:2008ig}
Fogli, G., Lisi, E., Marrone, A., {et~al.} 2008, Phys.Rev., D78, 033010

\bibitem[{Fogli {et~al.}(2011)Fogli, Lisi, Marrone, Palazzo, \&
  Rotunno}]{Fogli:2011qn}
Fogli, G., Lisi, E., Marrone, A., Palazzo, A., \& Rotunno, A. 2011

\bibitem[{{Fukugita} {et~al.}(1996){Fukugita}, {Ichikawa}, {Gunn}, {Doi},
  {Shimasaku}, \& {Schneider}}]{fukugitaetal96}
{Fukugita}, M., {Ichikawa}, T., {Gunn}, J.~E., {et~al.} 1996, \aj, 111, 1748

\bibitem[{Gomez-Cadenas {et~al.}(2011)Gomez-Cadenas, Martin-Albo, Mezzetto,
  Monrabal, \& Sorel}]{GomezCadenas:2011it}
Gomez-Cadenas, J., Martin-Albo, J., Mezzetto, M., Monrabal, F., \& Sorel, M.
  2011, * Temporary entry *

\bibitem[{Gonzalez-Garcia \& Maltoni(2008)}]{GonzalezGarcia:2007ib}
Gonzalez-Garcia, M., \& Maltoni, M. 2008, Phys.Rept., 460, 1

\bibitem[{Goobar {et~al.}(2006)Goobar, Hannestad, Mortsell, \&
  Tu}]{Goobar:2006xz}
Goobar, A., Hannestad, S., Mortsell, E., \& Tu, H. 2006, JCAP, 0606, 019

\bibitem[{{Gunn} {et~al.}(1998){Gunn}, {Carr}, {Rockosi}, {Sekiguchi}, {Berry},
  {Elms}, {de Haas}, {Ivezi{\'c}}, {Knapp}, {Lupton}, {Pauls}, {Simcoe},
  {Hirsch}, {Sanford}, {Wang}, {York}, {Harris}, {Annis}, {Bartozek},
  {Boroski}, {Bakken}, {Haldeman}, {Kent}, {Holm}, {Holmgren}, {Petravick},
  {Prosapio}, {Rechenmacher}, {Doi}, {Fukugita}, {Shimasaku}, {Okada}, {Hull},
  {Siegmund}, {Mannery}, {Blouke}, {Heidtman}, {Schneider}, {Lucinio}, \&
  {Brinkman}}]{gunnetal98}
{Gunn}, J.~E., {Carr}, M., {Rockosi}, C., {et~al.} 1998, \aj, 116, 3040

\bibitem[{{Gunn} {et~al.}(2006){Gunn}, {Siegmund}, {Mannery}, {Owen}, {Hull},
  {Leger}, {Carey}, {Knapp}, {York}, {Boroski}, {Kent}, {Lupton}, {Rockosi},
  {Evans}, {Waddell}, {Anderson}, {Annis}, {Barentine}, {Bartoszek}, {Bastian},
  {Bracker}, {Brewington}, {Briegel}, {Brinkmann}, {Brown}, {Carr},
  {Czarapata}, {Drennan}, {Dombeck}, {Federwitz}, {Gillespie}, {Gonzales},
  {Hansen}, {Harvanek}, {Hayes}, {Jordan}, {Kinney}, {Klaene}, {Kleinman},
  {Kron}, {Kresinski}, {Lee}, {Limmongkol}, {Lindenmeyer}, {Long}, {Loomis},
  {McGehee}, {Mantsch}, {Neilsen}, {Neswold}, {Newman}, {Nitta}, {Peoples},
  {Pier}, {Prieto}, {Prosapio}, {Rivetta}, {Schneider}, {Snedden}, \&
  {Wang}}]{gunnetal06}
{Gunn}, J.~E., {Siegmund}, W.~A., {Mannery}, E.~J., {et~al.} 2006, \aj, 131,
  2332

\bibitem[{{Guzik} {et~al.}(2007){Guzik}, {Bernstein}, \& {Smith}}]{guziketal07}
{Guzik}, J., {Bernstein}, G., \& {Smith}, R.~E. 2007, \mnras, 375, 1329

\bibitem[{{Hamann} {et~al.}(2008){Hamann}, {Hannestad}, {Melchiorri}, \&
  {Wong}}]{hamannetal08}
{Hamann}, J., {Hannestad}, S., {Melchiorri}, A., \& {Wong}, Y.~Y.~Y. 2008,
  \jcap, 7, 17

\bibitem[{{Hamaus} {et~al.}(2010){Hamaus}, {Seljak}, {Desjacques}, {Smith}, \&
  {Baldauf}}]{hamausetal10}
{Hamaus}, N., {Seljak}, U., {Desjacques}, V., {Smith}, R.~E., \& {Baldauf}, T.
  2010, \prd, 82, 043515

\bibitem[{Hannestad(2003)}]{Hannestad:2003xv}
Hannestad, S. 2003, JCAP, 0305, 004

\bibitem[{Hannestad(2005)}]{Hannestad:2005gj}
---. 2005, Phys.Rev.Lett., 95, 221301

\bibitem[{Hannestad \& Raffelt(2004)}]{Hannestad:2003ye}
Hannestad, S., \& Raffelt, G. 2004, JCAP, 0404, 008

\bibitem[{{Heavens} \& {Taylor}(1995)}]{heavenstaylor95}
{Heavens}, A.~F., \& {Taylor}, A.~N. 1995, \mnras, 275, 483

\bibitem[{{Ho} {et~al.}(2012){Ho}, {Cuesta}, \& {Seo}}]{hoetal11}
{Ho}, S., {Cuesta}, A., \& {Seo}, H.-J.~{\it et al}. 2012, submitted to ApJ

\bibitem[{Hu \& Eisenstein(1998)}]{Hu:1997vi}
Hu, W., \& Eisenstein, D.~J. 1998, Astrophys.J., 498, 497

\bibitem[{Jimenez {et~al.}(2010)Jimenez, Kitching, Pena-Garay, \&
  Verde}]{Jimenez:2010ev}
Jimenez, R., Kitching, T., Pena-Garay, C., \& Verde, L. 2010, JCAP, 1005, 035

\bibitem[{{Komatsu} \& {Seljak}(2002)}]{komsel02}
{Komatsu}, E., \& {Seljak}, U. 2002, \mnras, 336, 1256

\bibitem[{Komatsu {et~al.}(2009)}]{Komatsu:2008hk}
Komatsu, E., {et~al.} 2009, Astrophys.J.Suppl., 180, 330

\bibitem[{{Komatsu} {et~al.}(2010){Komatsu}, {Smith}, {Dunkley}, {Bennett},
  {Gold}, {Hinshaw}, {Jarosik}, {Larson}, {Nolta}, {Page}, {Spergel},
  {Halpern}, {Hill}, {Kogut}, {Limon}, {Meyer}, {Odegard}, {Tucker}, {Weiland},
  {Wollack}, \& {Wright}}]{Komatsuetal10}
{Komatsu}, E., {Smith}, K.~M., {Dunkley}, J., {et~al.} 2010

\bibitem[{Komatsu {et~al.}(2011)}]{Komatsu:2010fb}
Komatsu, E., {et~al.} 2011, Astrophys.J.Suppl., 192, 18

\bibitem[{{Larson} {et~al.}(2010){Larson}, {Dunkley}, {Hinshaw}, {Komatsu},
  {Nolta}, {Bennett}, {Gold}, {Halpern}, {Hill}, {Jarosik}, {Kogut}, {Limon},
  {Meyer}, {Odegard}, {Page}, {Smith}, {Spergel}, {Tucker}, {Weiland},
  {Wollack}, \& {Wright}}]{Larsonetal10}
{Larson}, D., {Dunkley}, J., {Hinshaw}, G., {et~al.} 2010, ArXiv e-prints

\bibitem[{Lesgourgues \& Pastor(2006)}]{Lesgourgues:2006nd}
Lesgourgues, J., \& Pastor, S. 2006, Phys.Rept., 429, 307

\bibitem[{{Lewis} \& {Bridle}(2002)}]{LewisBridle02}
{Lewis}, A., \& {Bridle}, S. 2002, Phys.Rev.D

\bibitem[{{Lewis} {et~al.}(2000){Lewis}, {Challinor}, \&
  {Lasenby}}]{LewChalLas00}
{Lewis}, A., {Challinor}, A., \& {Lasenby}, A. 2000, Astrophys.J.

\bibitem[{{Limber}(1954)}]{Limber:54}
{Limber}, D.~N. 1954, \apj, 119, 655

\bibitem[{Lobashev(2003)}]{Lobashev:2003kt}
Lobashev, V. 2003, Nucl.Phys., A719, 153

\bibitem[{{McDonald}(2006)}]{mcdonald06}
{McDonald}, P. 2006, \prd, 74, 103512

\bibitem[{Otten \& Weinheimer(2008)}]{Otten:2008zz}
Otten, E., \& Weinheimer, C. 2008, Rept.Prog.Phys., 71, 086201

\bibitem[{{Padmanabhan} {et~al.}(2003){Padmanabhan}, {Seljak}, \&
  {Pen}}]{padmaetal03}
{Padmanabhan}, N., {Seljak}, U., \& {Pen}, U.~L. 2003, \na, 8, 581

\bibitem[{{Padmanabhan} {et~al.}(2007){Padmanabhan}, {Schlegel}, {Seljak},
  {Makarov}, {Bahcall}, {Blanton}, {Brinkmann}, {Eisenstein}, {Finkbeiner},
  {Gunn}, {Hogg}, {Ivezi{\'c}}, {Knapp}, {Loveday}, {Lupton}, {Nichol},
  {Schneider}, {Strauss}, {Tegmark}, \& {York}}]{padmaetal07}
{Padmanabhan}, N., {Schlegel}, D.~J., {Seljak}, U., {et~al.} 2007, \mnras, 378,
  852

\bibitem[{{Percival} {et~al.}(2010){Percival}, {Reid}, {Eisenstein}, {Bahcall},
  {Budavari}, {Frieman}, {Fukugita}, {Gunn}, {Ivezi{\'c}}, {Knapp}, {Kron},
  {Loveday}, {Lupton}, {McKay}, {Meiksin}, {Nichol}, {Pope}, {Schlegel},
  {Schneider}, {Spergel}, {Stoughton}, {Strauss}, {Szalay}, {Tegmark},
  {Vogeley}, {Weinberg}, {York}, \& {Zehavi}}]{Percetal10}
{Percival}, W.~J., {Reid}, B.~A., {Eisenstein}, D.~J., {et~al.} 2010, MNRAS

\bibitem[{{Pier} {et~al.}(2003){Pier}, {Munn}, {Hindsley}, {Hennessy}, {Kent},
  {Lupton}, \& {Ivezi{\'c}}}]{pieretal03}
{Pier}, J.~R., {Munn}, J.~A., {Hindsley}, R.~B., {et~al.} 2003, \aj, 125, 1559

\bibitem[{Reid {et~al.}(2010)Reid, Percival, Eisenstein, Verde, Spergel,
  {et~al.}}]{Reid:2009xm}
Reid, B.~A., Percival, W.~J., Eisenstein, D.~J., {et~al.} 2010,
  Mon.Not.Roy.Astron.Soc., 404, 60, * Brief entry *

\bibitem[{{Reid} {et~al.}(2010{\natexlab{a}}){Reid}, {Verde}, {Jimenez}, \&
  {Mena}}]{Reid:2009nq}
{Reid}, B.~A., {Verde}, L., {Jimenez}, R., \& {Mena}, O. 2010{\natexlab{a}},
  \jcap, 1, 3

\bibitem[{{Reid} {et~al.}(2010{\natexlab{b}}){Reid}, {Verde}, {Jimenez}, \&
  {Mena}}]{reidetal10}
---. 2010{\natexlab{b}}, 1, 3

\bibitem[{{Riemer--S{\o}rensen} {et~al.}(2011){Riemer--S{\o}rensen}, {Blake},
  {Parkinson}, {Davis}, {Brough}, {Colless}, {Contreras}, {Couch}, {Croom},
  {Croton}, {Drinkwater}, {Forster}, {Gilbank}, {Gladders}, {Glazebrook},
  {Jelliffe}, {Jurek}, {Li}, {Madore}, {Martin}, {Pimbblet}, {Poole}, {Pracy},
  {Sharp}, {Wisnioski}, {Woods}, {Wyder}, \& {Yee}}]{riemeretal11}
{Riemer--S{\o}rensen}, S., {Blake}, C., {Parkinson}, D., {et~al.} 2011, ArXiv
  e-prints

\bibitem[{{Riess} {et~al.}(2011){Riess}, {Macri}, {Casertano}, {Lampeitl},
  {Ferguson}, {Filippenko}, {Jha}, {Li}, \& {Chornock}}]{hst11}
{Riess}, A.~G., {Macri}, L., {Casertano}, S., {et~al.} 2011, \apj, 730, 119

\bibitem[{{Ross} {et~al.}(2011){Ross}, {Ho}, {Cuesta}, {Tojeiro}, {Percival},
  {Wake}, {Masters}, {Nichol}, {Myers}, {de Simoni}, {Seo},
  {Hern{\'a}ndez-Monteagudo}, {Crittenden}, {Blanton}, {Brinkmann}, {da Costa},
  {Guo}, {Kazin}, {Maia}, {Maraston}, {Padmanabhan}, {Prada}, {Ramos},
  {Sanchez}, {Schlafly}, {Schlegel}, {Schneider}, {Skibba}, {Thomas}, {Weaver},
  {White}, \& {Zehavi}}]{rossetal11}
{Ross}, A.~J., {Ho}, S., {Cuesta}, A.~J., {et~al.} 2011, \mnras, 417, 1350

\bibitem[{{Saito} {et~al.}(2008){Saito}, {Takada}, \& {Taruya}}]{saitoetal08}
{Saito}, S., {Takada}, M., \& {Taruya}, A. 2008, Physical Review Letters, 100,
  191301

\bibitem[{{Saito} {et~al.}(2009){Saito}, {Takada}, \& {Taruya}}]{saitoetal09}
---. 2009, \prd, 80, 083528

\bibitem[{{Saito} {et~al.}(2011){Saito}, {Takada}, \& {Taruya}}]{saitoetal11}
---. 2011, \prd, 83, 043529

\bibitem[{{Scherrer} \& {Weinberg}(1998)}]{scherrwein98}
{Scherrer}, R.~J., \& {Weinberg}, D.~H. 1998, \apj, 504, 607

\bibitem[{{Schulz} \& {White}(2006)}]{schulzwhite06}
{Schulz}, A.~E., \& {White}, M. 2006, Astroparticle Physics, 25, 172

\bibitem[{Schwetz {et~al.}(2011)Schwetz, Tortola, \& Valle}]{Schwetz:2011zk}
Schwetz, T., Tortola, M., \& Valle, J. 2011

\bibitem[{{Seljak}(1998)}]{seljak98}
{Seljak}, U. 1998, \apj, 506, 64

\bibitem[{{Seljak}(2000)}]{seljak2000}
---. 2000, \mnras, 318, 203

\bibitem[{{Seljak}(2001)}]{seljak01}
---. 2001, \mnras, 325, 1359

\bibitem[{Seljak {et~al.}(2006)Seljak, Slosar, \& McDonald}]{Seljak:2006bg}
Seljak, U., Slosar, A., \& McDonald, P. 2006, JCAP, 0610, 014

\bibitem[{Seljak {et~al.}(2005)}]{Seljak:2004xh}
Seljak, U., {et~al.} 2005, Phys.Rev., D71, 103515, <a eprint =

\bibitem[{{Seo} {et~al.}(2012){Seo}, {Ho}, \& {White}}]{seobao}
{Seo}, H.-J., {Ho}, S., \& {White}, M.~{\it et al}. 2012, submitted to ApJ

\bibitem[{{Smith} {et~al.}(2003){Smith}, {Peacock}, {Jenkins}, {White},
  {Frenk}, {Pearce}, {Thomas}, {Efstathiou}, \& {Couchman}}]{Smithetal03}
{Smith}, R.~E., {Peacock}, J.~A., {Jenkins}, A., {et~al.} 2003, \mnras, 341,
  1311

\bibitem[{Spergel {et~al.}(2003)}]{Spergel:2003cb}
Spergel, D., {et~al.} 2003, Astrophys.J.Suppl., 148, 175

\bibitem[{Spergel {et~al.}(2007)}]{Spergel:2006hy}
---. 2007, Astrophys.J.Suppl., 170, 377

\bibitem[{{Swanson} {et~al.}(2010){Swanson}, {Percival}, \&
  {Lahav}}]{swansonetal10}
{Swanson}, M.~E.~C., {Percival}, W.~J., \& {Lahav}, O. 2010, \mnras, 409, 1100

\bibitem[{{Tegmark} {et~al.}(1998){Tegmark}, {Hamilton}, {Strauss}, {Vogeley},
  \& {Szalay}}]{tegmarketal98}
{Tegmark}, M., {Hamilton}, A.~J.~S., {Strauss}, M.~A., {Vogeley}, M.~S., \&
  {Szalay}, A.~S. 1998, \apj, 499, 555

\bibitem[{Tegmark {et~al.}(2004)}]{Tegmark:2003ud}
Tegmark, M., {et~al.} 2004, Phys.Rev., D69, 103501

\bibitem[{Thomas {et~al.}(2010)Thomas, Abdalla, \& Lahav}]{Thomas:2009ae}
Thomas, S.~A., Abdalla, F.~B., \& Lahav, O. 2010, Phys.Rev.Lett., 105, 031301

\bibitem[{{White} {et~al.}(2011){White}, {Blanton}, {Bolton}, {Schlegel},
  {Tinker}, {Berlind}, {da Costa}, {Kazin}, {Lin}, {Maia}, {McBride},
  {Padmanabhan}, {Parejko}, {Percival}, {Prada}, {Ramos}, {Sheldon}, {de
  Simoni}, {Skibba}, {Thomas}, {Wake}, {Zehavi}, {Zheng}, {Nichol},
  {Schneider}, {Strauss}, {Weaver}, \& {Weinberg}}]{whiteetal11}
{White}, M., {Blanton}, M., {Bolton}, A., {et~al.} 2011, \apj, 728, 126

\bibitem[{{York} {et~al.}(2000){York}, {Adelman}, {Anderson}, {Anderson},
  {Annis}, {Bahcall}, {Bakken}, {Barkhouser}, {Bastian}, {Berman}, {Boroski},
  {Bracker}, {Briegel}, {Briggs}, {Brinkmann}, {Brunner}, {Burles}, {Carey},
  {Carr}, {Castander}, {Chen}, {Colestock}, {Connolly}, {Crocker}, {Csabai},
  {Czarapata}, {Davis}, {Doi}, {Dombeck}, {Eisenstein}, {Ellman}, {Elms},
  {Evans}, {Fan}, {Federwitz}, {Fiscelli}, {Friedman}, {Frieman}, {Fukugita},
  {Gillespie}, {Gunn}, {Gurbani}, {de Haas}, {Haldeman}, {Harris}, {Hayes},
  {Heckman}, {Hennessy}, {Hindsley}, {Holm}, {Holmgren}, {Huang}, {Hull},
  {Husby}, {Ichikawa}, {Ichikawa}, {Ivezi{\'c}}, {Kent}, {Kim}, {Kinney},
  {Klaene}, {Kleinman}, {Kleinman}, {Knapp}, {Korienek}, {Kron}, {Kunszt},
  {Lamb}, {Lee}, {Leger}, {Limmongkol}, {Lindenmeyer}, {Long}, {Loomis},
  {Loveday}, {Lucinio}, {Lupton}, {MacKinnon}, {Mannery}, {Mantsch}, {Margon},
  {McGehee}, {McKay}, {Meiksin}, {Merelli}, {Monet}, {Munn}, {Narayanan},
  {Nash}, {Neilsen}, {Neswold}, {Newberg}, {Nichol}, {Nicinski}, {Nonino},
  {Okada}, {Okamura}, {Ostriker}, {Owen}, {Pauls}, {Peoples}, {Peterson},
  {Petravick}, {Pier}, {Pope}, {Pordes}, {Prosapio}, {Rechenmacher}, {Quinn},
  {Richards}, {Richmond}, {Rivetta}, {Rockosi}, {Ruthmansdorfer}, {Sandford},
  {Schlegel}, {Schneider}, {Sekiguchi}, {Sergey}, {Shimasaku}, {Siegmund},
  {Smee}, {Smith}, {Snedden}, {Stone}, {Stoughton}, {Strauss}, {Stubbs},
  {SubbaRao}, {Szalay}, {Szapudi}, {Szokoly}, {Thakar}, {Tremonti}, {Tucker},
  {Uomoto}, {Vanden Berk}, {Vogeley}, {Waddell}, {Wang}, {Watanabe},
  {Weinberg}, {Yanny}, \& {Yasuda}}]{SDSS}
{York}, D.~G., {Adelman}, J., {Anderson}, Jr., J.~E., {et~al.} 2000, \apj, 120,
  1579

\end{thebibliography}

\end{document}